\UseRawInputEncoding
\documentclass[%
10pt,
twocolumn,
 amsmath,amssymb,
 aps,
 superscriptaddress,
pra,
floatfix
]{revtex4-2}
\usepackage{graphicx}
\usepackage{dcolumn}  
\usepackage{bm}       
\usepackage{svg}      
\usepackage{upgreek}  
\usepackage{amssymb}  
\usepackage{makecell} 
\usepackage{xcolor}   
\usepackage[normalem]{ulem} 
\usepackage[official]{eurosym}
\usepackage{todonotes} 
\usepackage{braket}   
\usepackage{url}      
\usepackage{lipsum}
\usepackage{multirow}
\usepackage{enumerate} 
\usepackage[hidelinks]{hyperref}
\usepackage[capitalise]{cleveref} 

\definecolor{DarkGreen}{rgb}{0, 0.5, 0}

\usepackage{soul}
\setstcolor{red} 

\definecolor{brightube}{rgb}{0.82, 0.62, 0.91}

\makeatletter
\newsavebox{\@brx}
\newcommand{\llangle}[1][]{\savebox{\@brx}{\(\m@th{#1\langle}\)}%
  \mathopen{\copy\@brx\kern-0.5\wd\@brx\usebox{\@brx}}}
\newcommand{\rrangle}[1][]{\savebox{\@brx}{\(\m@th{#1\rangle}\)}%
  \mathclose{\copy\@brx\kern-0.5\wd\@brx\usebox{\@brx}}}
\makeatother

\newboolean{showkeys}
\setboolean{showkeys}{false}
\ifthenelse{\boolean{showkeys}}{\usepackage[notref,notcite]{showkeys}} 
{}

\usepackage{scalerel}
\usepackage{stackengine}

\newcommand{\approptoinn}[2]{\mathrel{\vcenter{
  \offinterlineskip\halign{\hfil$##$\cr
    #1\propto\cr\noalign{\kern2pt}#1\sim\cr\noalign{\kern-2pt}}}}}

\begin{document}

\title{Variational Quantum Eigensolver for Real-World Finance: Scalable Solutions for Dynamic Portfolio Optimization Problems}


\author{I. De Le\'on}
\affiliation{Global Data Quantum, Gran V\'ia de Don Diego L\'opez de Haro, 1, 48001 Bilbo, Bizkaia, Spain}
\author{D. Arias}
\affiliation{Global Data Quantum, Gran V\'ia de Don Diego L\'opez de Haro, 1, 48001 Bilbo, Bizkaia, Spain}
\affiliation{Universidad de Deusto, Avda. de las Universidades, 24, 48007, Bilbo, Bizkaia, Spain}
\author{M. Mart\'in-Cordero}
\affiliation{Global Data Quantum, Gran V\'ia de Don Diego L\'opez de Haro, 1, 48001 Bilbo, Bizkaia, Spain}
\author{M. E. Molina}
\affiliation{Global Data Quantum, Gran V\'ia de Don Diego L\'opez de Haro, 1, 48001 Bilbo, Bizkaia, Spain}
\author{P. Serrano}
\affiliation{BBVA Quantum, Calle Azul 4, 28050, Madrid, Spain}
\author{S. Hern\'andez-Santana}
\affiliation{BBVA Quantum, Calle Azul 4, 28050, Madrid, Spain}
\author{M. A. J. Herrera}
\affiliation{Global Data Quantum, Gran V\'ia de Don Diego L\'opez de Haro, 1, 48001 Bilbo, Bizkaia, Spain}
\author{J. Fraxanet}
\affiliation{IBM Quantum, IBM Thomas J Watson Research Center, Yorktown Heights, NY 10598, USA}
\author{G. Carrascal}
\affiliation{IBM Quantum, IBM Spain, Plaza Pablo Ruiz Picasso 11, 28020, Madrid, Spain}
\author{E. S\'anchez }
\affiliation{BBVA Quantum, Calle Azul 4, 28050, Madrid, Spain}
\author{I. Posadillo}
\affiliation{Global Data Quantum, Gran V\'ia de Don Diego L\'opez de Haro, 1, 48001 Bilbo, Bizkaia, Spain}
\author{\'A. Nodar}
\affiliation{Global Data Quantum, Gran V\'ia de Don Diego L\'opez de Haro, 1, 48001 Bilbo, Bizkaia, Spain}

\date{\today}
\begin{abstract}

We present a scalable, hardware-aware methodology for extending the Variational Quantum Eigensolver (VQE) to large, realistic Dynamic Portfolio Optimization (DPO) problems. Building on the scaling strategy from our previous work~\cite{nodar2024scaling}, where we tailored a VQE workflow to both the DPO formulation and the target QPU, we now put forward two significant advances. The first is the implementation of the Ising Sample-based Quantum Configuration Recovery (ISQR) routine, which improves solution quality in Quadratic Unconstrained Binary Optimization problems. The second is the use of the VQE Constrained method~\cite{mugel2022dynamic} to decompose the optimization task, enabling us to handle DPO instances with more variables than the available qubits on current hardware. These advances, which are broadly applicable to other optimization problems, allow us to address a portfolio with a size relevant to the financial industry, consisting of up to 38 assets and covering the full Spanish stock index (IBEX 35). Our results, obtained on a real Quantum Processing Unit (IBM Fez), show that this tailored workflow achieves financial performance on par with classical methods while delivering a broader set of high-quality investment strategies, demonstrating a viable path towards obtaining practical advantage from quantum optimization in real financial applications.

\end{abstract}

\maketitle

\section{Introduction} \label{sec:Intro}

Advances in algorithm design are driving quantum computing towards efficiency, cost-effectiveness, and accuracy over classical algorithms~\cite{lanes2025framework}. In particular, recent research on quantum optimization at utility scale~\cite{romero2024bias,montanez2024towards,pelofske2024scaling,lanes2025framework,sachdeva2024quantum} is narrowing the gap to quantum advantage. The hardware development of gate-based Quantum Processing Units (QPUs) has led to the demonstration of quantum utility~\cite{kim2023evidence, montanez2025evaluating} in different fields, such as chemistry~\cite{robledo2024chemistry, danilov2025enhancing,kaliakin2025accurate,barison2025quantum,shajan2024toward}, quantum simulation~\cite{yu2023simulating,kim2023evidence,farrell2024quantum,shinjo2024unveiling,cobos2025real,switzer2025realization}, and machine learning~\cite{park2020practical,heredge2021quantum,larose2020robust,hur2022quantum,beer2020training}. Quantum optimization has also made its way into the area of finance. The Dynamic Portfolio Optimization (DPO) problem focuses on rebalancing investment allocations over time to maximize investment return while minimizing risks, taking into account evolving market conditions and investor preferences. The complexity of this problem scales rapidly with the number of variables involved, leading to significant challenges for classical approaches and making it a highlighted candidate for a potential quantum advantage~\cite{mugel2022dynamic, buonaiuto2023best, brandhofer2023benchmarking, carrascal2023backtesting, nodar2024scaling}.

In this work, we demonstrate the performance of quantum optimizers on real QPUs for large-scale portfolio optimization. Our study focuses on portfolio sizes that reflect realistic industry conditions, encompassing up to 38 assets that include the complete Spanish stock index (IBEX 35) along with additional region-specific assets (see Appendix~\ref{app:data_prep}). We optimize investment allocations over temporal windows in which portfolio returns are comparable to transaction costs, thereby introducing challenging trading dynamics. Taken together, the scale and composition of our instance closely mirror real-world financial scenarios, underscoring the current practical relevance and impact of quantum solutions for portfolio optimization.

Addressing these real-world problems presents two main challenges: i) managing large portfolios, which implies increasing the number of variables required to solve the DPO problem, and ii) enhancing the performance of previously proposed quantum optimization methods~\cite{nodar2024scaling, mugel2022dynamic}, making them competitive with classical optimizers. In this context, hybrid quantum algorithms are the leading method for addressing complex optimization problems on current gate-based QPUs~\cite{kim2023evidence}. These hybrid approaches integrate classical and quantum techniques, with the Variational Quantum Eigensolver (VQE) results~\cite{tilly2022variational, carrascal2024differential, LIU2024102117, cerezo2021variational} as a prime example of their efficacy. Specifically, here we build upon our previous VQE workflow in Ref.~\cite{nodar2024scaling}, which introduced a quantum optimization framework for large-scale DPO problems. As an extension to this approach, we present a noise-aware post-processing technique based on the Sample-based Quantum Diagonalization (SQD) algorithm~\cite{robledo2024chemistry, barison2025quantum} and implement the problem-division strategy from Ref.~\cite{mugel2022dynamic} to enable scaling to larger portfolios.

To enhance the performance and consistency of the quantum optimizer, we adapt the SQD post-processing technique~\cite{robledo2024chemistry} to the DPO problem. Specifically, we adapt the Configuration Recovery (CR) subroutine from SQD, which probabilistically corrects bit-flip errors through iterative classical post-processing. We refer to this adaptation of the SQD as Ising Sample-based Quantum configuration Recovery (ISQR), since the quantum formulation of the DPO problem yields an already diagonal Ising Hamiltonian. ISQR enforces the penalty factors in the post-processing stage by incorporating them as symmetries through CR, using the structure of the investment constraints to guide the recovery process and improve the consistency of solutions. To the best of our knowledge, this is one of the first works to extend an SQD-inspired technique beyond quantum chemistry and materials science, highlighting its broad applicability and impact.

\begin{figure}[t!]
\includegraphics[width=0.49\textwidth]{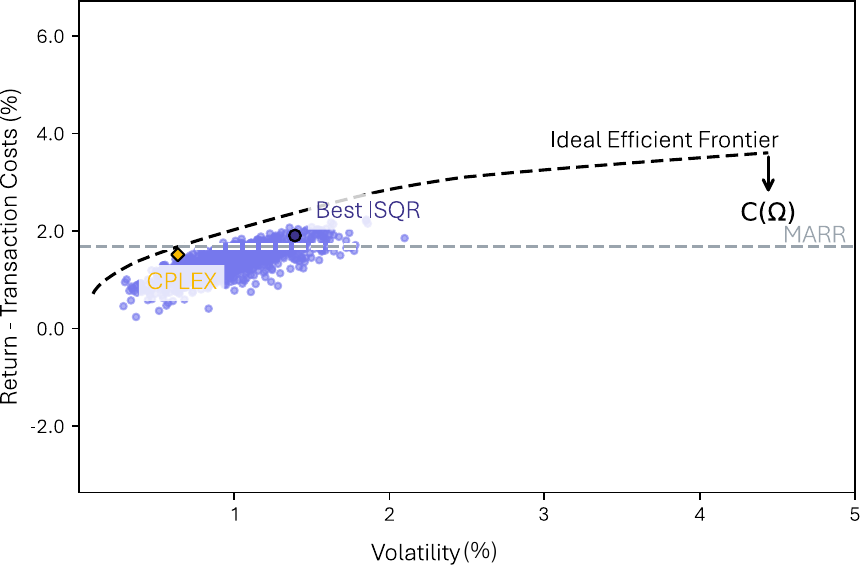}
\caption{Financial performance of the ISQR and CPLEX solutions for a 9-asset DPO problem in a fixed investment window ($t=4$, see Section~\ref{sec:944}). The horizontal axis represents investment volatility (associated to the risk), while the vertical axis shows the effective return, defined as the direct return minus transaction costs. We indicate in the dashed gray line our reference MARR of $1.68\%$. The dashed black curve indicates the \textit{ideal} efficient frontier obtained with PyPortfolioOptimizer~\cite{Martin2021}, representing the optimal trade-off between return and volatility without accounting for transaction costs--thus, it is not achievable in practice. Blue dots correspond to a selection of the 1000 optimized investments from the ISQR method with lowest (best) optimization cost. The dark blue dot marks the single ISQR solution with the overall lowest (best) optimization cost, while the yellow diamond represents the CPLEX solution.\label{fig:fig_intro}}
\end{figure}

Figure~\ref{fig:fig_intro} shows the ISQR post-processed solutions (blue dots) obtained for a 9-asset problem at a specific investment window (see caption of Fig.~\ref{fig:fig_intro}, and section~\ref{sec:944}, below). The dark blue dot represents the best-performing ISQR post-processed solution in terms of the optimization objective function. To frame the financial performance of these results, we compare them against an \emph{ideal} efficient frontier (dashed black curve in the figure), which defines the \emph{ideal} investment with the lowest risk for a given expected return, without transaction costs. We also include a Minimum Acceptable Rate of Return (MARR) of $1.68\%$ (dashed gray line in the figure)~\cite{ECB}, which indicates investments with a desired annual return of $22.04\%$ (twice the historical return rate of the S\&P500, see Section~\ref{sec:Efficient_Frontier}). Due to conservative investment preferences, our results cluster on the left-side of the graph, with low volatilities and very close to the ideal efficient frontier. Most importantly, some of the ISQR solutions, and specifically, the best performing solution (dark blue dot), surpass the MARR threshold. In contrast, the solution obtained with the IBM Decision Optimization CPLEX~\cite{docplex} classical optimizer (yellow diamond) does not surpass the MARR threshold, resulting in a worse financial evaluation. Section~\ref{sec:944} provides a deeper discussion of these results.

In parallel with the ISQR routine, we implement the VQE Constrained (VQEC) approach from Ref.~\cite{mugel2022dynamic} to tackle larger DPO problems. The VQEC decomposes the full DPO problem to optimize the portfolio step by step on the QPU. This decomposition  significantly increases the number of assets involved, enabling the analysis of portfolios relevant to the financial industry. Integrating ISQR into both VQE and VQEC further improves the consistency and quality of the solutions. Together, these techniques produce optimized investment strategies comparable to those obtained with state-of-the-art classical methods~\cite{docplex} while extending the scalability of quantum optimization. As quantum hardware continues to develop, this methodology could serve as a foundation for solving large-scale portfolio problems beyond classical computational limits.

This work is structured as follows: Section~\ref{sec:DPortfolio} presents the formulation of the DPO problem into an Ising Hamiltonian, as well as the metrics we use to evaluate the results. Section~\ref{sec:QuantumSolution} reviews the VQE algorithm, including the techniques we use to accelerate it, and presents the ISQR routine. Section~\ref{sec:944} shows the results for a 9-asset DPO problem, using the ISQR post-processing routine. Section~\ref{sec:ExploringVQEC} introduces the VQEC approach and its application to a 9- and 38-asset DPO problems, where we emphasize the scalability aspect of the VQEC strategy.

\section{Dynamic Portfolio Optimization}
\label{sec:DPortfolio}
We first introduce the formulation for the DPO problem in Subsection~\ref{sec:Formulation}. In Subsection~\ref{sec:Efficient_Frontier}, we present the metrics used to study the quality of the solutions obtained. 

\subsection{Formulation of the Dynamic Portfolio Optimization problem}
\label{sec:Formulation}

The objective of the DPO problem is to find the best investment strategy $\bm{\Omega}$ that maximizes the return of the investment, $F(\bm{\Omega})$, while minimizing the risk, $R(\bm{\Omega})$, transaction costs, $C(\bm{\Omega})$, and complying with investor restrictions, $\Gamma(\bm{\Omega})$. Here, $\bm{\Omega}= \{\bm{\omega}_{1}, \dots , \bm{\omega}_{N_t}\}$ represents the complete strategy for investment rebalance across the portfolio over $N_t$ time steps. For a portfolio with $N_a$ assets, the distribution of the investment at a fixed time $t$ is defined as $\bm{\omega}_t = (\omega_{t, 1}, \dots, \omega_{t, N_a})$. Hence, an investment strategy is described by a total of $N_a\times N_t$ weights, where each ${\omega}_{t,a}$ designates the decimal percentage of the investment at time $t$ in the asset $a$. The formulation used in this paper for the DPO problem builds upon the formulation in Refs.~\cite{nodar2024scaling, mugel2022dynamic}, where the DPO problem is expressed as a minimization Quadratic Unconstrained Binary Optimization (QUBO) objective function:
\begin{equation}
    \label{eq:QUBO}
    Q(\bm{\Omega}) = -F(\mathbf{\Omega}) +\frac{\gamma}{2}R(\mathbf{\Omega}) +C(\mathbf{\Omega}) + \Gamma(\mathbf{\Omega}).
\end{equation}
In Eq.~\eqref{eq:QUBO} above, the expected return is given by 
\begin{equation}
    F(\bm{\Omega}) = \sum_{t=1}^{N_{t}} \bm{\mu}_t^T\bm{\omega}_t,
    \label{eq:return}
\end{equation}
with $\bm{\mu}_t = (\mu_{t,1}, \dots, \mu_{t, N_a})$ denoting the return vector. Each $\mu_{t, a}$ represents the logarithmic return comparing the price of asset $a$ at time $t$, $P_{t,a}$, with its price at time $t-1$, $P_{t-1,a}$,
\begin{equation}
    \mu_{t, a}=\text{log}\left(\frac{P_{t,a}}{P_{t-1,a}}\right).
    \label{eq:mu_ta}
\end{equation}
The risk is given by
\begin{equation}
    R(\bm{\Omega}) =  \sum_{t =1}^{N_t} \bm{\omega}_t^T \Sigma_t \bm{\omega}_t,
    \label{eq:risk}
\end{equation}
where $\Sigma_t$ is the covariance matrix of the logarithmic return calculated for each time step  -- see Eq.~\eqref{eq:covariance_matrix} in Appendix~\ref{app:subapp:covariance_matrix}. In Eq.~\eqref{eq:QUBO}, the risk is weighted by a factor $\gamma/2$~\cite{markowits1952portfolio}, with $\gamma$ being the risk aversion coefficient (see Appendix~\ref{app:data_prep}).

The definition of the transaction costs $C(\bm{\Omega})$ in this work differs from the formulation introduced in Ref.~\cite{nodar2024scaling}. To account for price changes when liquidating a position, we adjust the strategy from the previous time step according to the price dynamics of each asset, ($P_{t,a}/P_{t-1,a}$), corrected by an estimated portfolio growth factor $g = (1+\epsilon)^{\Delta t /365}$. The parameter $\Delta t$ is the number of days between investment rebalances (set in this work to $\Delta t = 28$ days), while $\epsilon$ is the annual growth rate~\footnote{In this work, we use $\epsilon=0.18$ as an estimation of the annual growth rate during the time period considered.}. Defining $\varphi_{t,a}=P_{t,a}/(gP_{t-1,a})$ as an estimate of how the portfolio weights evolve between investment rebalances, the final expression for the transaction costs follows:
\begin{equation}
    \label{eq:transaction_costs}
    C(\bm{\Omega})= \sum_{t=1}^{N_t } \sum_{a=1}^{N_a} \nu_a\lambda_{t,a}\left(\omega_{t, a} -\varphi_{t,a}\omega_{t-1,a}\right)^2.
\end{equation}
The parameter $\nu_a$ denotes the asset-specific transaction fee (Table~\ref{tab:tickers_param} shows the values of $\nu_a$ used in this work). Next, the coefficient $\lambda_{t,a}$ is a piecewise constant function that allows us to write the DPO problem in the QUBO framework -- more details in Appendix~\ref{app:trans_costs}. Last, for the first time step, we define the initial investment in the portfolio as $\bm{\omega}_{0}=\bm{0}$ for all cases studied in this work.

As mentioned before, each $\omega_{t,a}$ represents the fraction of the portfolio invested in asset $a$ at time $t$. Hence, $\omega_{t,a}$ must satisfy:
\begin{equation}
    \label{eq:constraint_original_problem}
    \sum_{a=1}^{N_a} \omega_{t,a} = 1, \qquad \forall t,
\end{equation}
that is, the total allocation cannot exceed the total amount available for investment. We introduce the restriction given by Eq.~\eqref{eq:constraint_original_problem} in the objective function in Eq.~\eqref{eq:QUBO} as a quadratic penalty term modulated by a Lagrange multiplier $\rho$~\cite{mugel2022dynamic, nodar2024scaling} (see Appendix~\ref{app:data_prep}):
\begin{equation}
    \Gamma(\mathbf{\Omega}) = \rho \sum_{t=1}^{N_t }\left( \sum_{a=1}^{N_a} \omega_{t,a} -1\right)^2.
    \label{eq:constraint_gamma}
\end{equation}
Note that, in this work, we reserve the term ``restrictions'' for $\Gamma(\bm{\Omega})$ instead of the usual ``constraints'', to avoid confusion over the VQEC~\cite{mugel2022dynamic} methodology in Section~\ref{sec:ExploringVQEC}. 

Next, in order to frame the objective function in Eq.~\eqref{eq:QUBO} into the QUBO formalism, we write $\omega_{t,a}$ in terms of binary variables $x_{t,a,r}$:
\begin{equation}
    \omega_{t,a} = m_a + \frac{(B_a-m_a)}{2^{N_r}-1} \sum_{r=1}^{N_r} 2^{r-1} x_{t,a,r},
    \label{eq:binary_conversion}
\end{equation}
with subscripts $t$ and $a$ representing the time index and the asset index, respectively. Subscript $r$ stands for the bit resolution index and ranges from $1$ to $N_r$, where $N_r$ is the number of binary variables used to describe the investment $\omega_{t,a}$, for each time and asset. In Eq.~\eqref{eq:binary_conversion}, $B_a\in[0,1]$ and $m_a\in[0,1]$ are the maximum and minimum investment per asset respectively, whose values can be found in Table~\ref{tab:tickers_param}. This encoding entails:
\begin{equation}
    \label{eq:new_constraint}
    m_a\leq\omega_{t,a}\leq B_a.
\end{equation}
We can only guarantee that Eq.~\eqref{eq:new_constraint} is fulfilled if Eq.~\eqref{eq:constraint_original_problem} is satisfied. However, the opposite (Eq.~\eqref{eq:constraint_original_problem} is fulfilled if Eq.~\eqref{eq:new_constraint} is satisfied) is not necessarily true, due to the $\omega_{t,a}$ definition in Eq.~\eqref{eq:binary_conversion}. Furthermore, we note that the restrictions imposed in Eq.~\eqref{eq:constraint_original_problem} are not strict, since QUBO problems are unconstrained by definition, which allows slight deviations from restrictions to be tolerated in the final solution. In that case, further renormalization must be done since it is not possible to allocate more than $100\%$ of the investment. For this purpose, we follow a normalization procedure to ensure both the restriction in Eq.~\eqref{eq:constraint_original_problem} and the compliance with the investment bounds in Eq.~\eqref{eq:new_constraint} (our normalization procedure is described in Appendix~\ref{app:normalization}).

Last, we map our QUBO problem to an Ising Hamiltonian, $H_\text{Ising}$,by expressing the $x_{t,a,r}$ variables (Eq.~\eqref{eq:binary_conversion}) as Pauli operators $Z_q$~\cite{nodar2024scaling}. We perform the change of variables $x_{t,a,r}\to (1-Z_q)/2$. The subscript $q$ refers to the $q$-th logical qubit and it is defined as:
\begin{equation}
    q=r+N_r\times a+ t\times(N_a\times N_r),
    \label{eq:q_def}
\end{equation}
with $q\in\left[0, N_q-1\right]$, being $N_q$ the total number of logical qubits needed to solve the QUBO problem:
\begin{equation}
    N_q=N_a\times N_r\times N_t.
    \label{eq:Nq_def}
\end{equation}

In this work, we use this formulation to solve the DPO problem with VQE for portfolios of $N_a = 9$ assets (Section~\ref{sec:944}), and with VQEC for both $N_a = 9$ and $N_a = 38$ assets (Sections~\ref{sec:94t} and ~\ref{sec:384t}, respectively). In all cases, we use a fixed resolution of $N_r = 4$ bits and $N_t = 4$ time steps.

\subsection{Financial scores}
\label{sec:Efficient_Frontier}

The portfolio optimization cost is directly computed from Eq.~\eqref{eq:QUBO} in Section~\ref{sec:Formulation}, which serves as the objective function guiding our optimization, with lower values indicating a more effective investment strategy. However, we also evaluate the financial performance of the solution under four additional scores: Sharpe ratio, volatility, effective return, and proximity to the efficient frontier.

The Sharpe ratio is a widely used score in financial analysis first introduced in Ref.~\cite{sharpe1966mutual}. In particular, here we use the annualized Sharpe ratio~\cite{special_sharpe},
\begin{equation}
    S(\bm{\Omega}) = \frac{\text{Ann. Eff. Return} - \text{RFR}}{\text{Ann. Volatility}},
    \label{eq:sharpe_ratio}
\end{equation}
where the annualized volatility in the denominator is calculated as
\begin{equation}
    \text{Ann. Volatility} = \sqrt{\frac{365}{\Delta t \times N_t}} \sqrt{\Delta t \times R(\bm{\Omega})},
    \label{eq:ann_volatility}
\end{equation}
with $\sqrt{\Delta t \times R(\bm{\Omega})}$ being the volatility over the $N_t$ periods of $\Delta t$ days.

$\text{RFR}$ in Eq.~\eqref{eq:sharpe_ratio} denotes the risk-free rate, which we take as $\text{RFR} = 1.72\%$, corresponding to the average Euro short-term rate over the considered period~\cite{ECB}. The annualized effective return is defined as
\begin{equation}
    \text{Ann. Eff. Return} = \left[ F(\boldsymbol{\Omega}) - C (\boldsymbol{\Omega}) \right] \frac{365}{N_t \times \Delta_t},
    \label{eq:ann_eff_return}
\end{equation}
where $F(\boldsymbol{\Omega}) - C(\boldsymbol{\Omega})$ represents the effective return over a horizon of $N_t \times \Delta_t$ days.  

In this work, the annualized effective return is used as a primary performance metric. Further, we reflect an ambitious investor requiring an annualized Minimum Acceptable Rate of Return (MARR) corresponding to twice the average annual return of the S\&P~500 index over the ten-year period ending in 2023 (the final year of the dataset considered in this work). This average annual return is $11.02\%$, which yields an annualized MARR of $22.04\%$, and corresponds to a rate of
\begin{equation}
    \text{MARR} = \frac{22.04\%}{365/\Delta t} \approx 1.68\%,
\end{equation}
over each $\Delta t$ interval, which we also adopt as a benchmark for evaluating investment performance at rebalancing time $t$.

\begin{figure}[t!]
\includegraphics[width=0.45\textwidth]{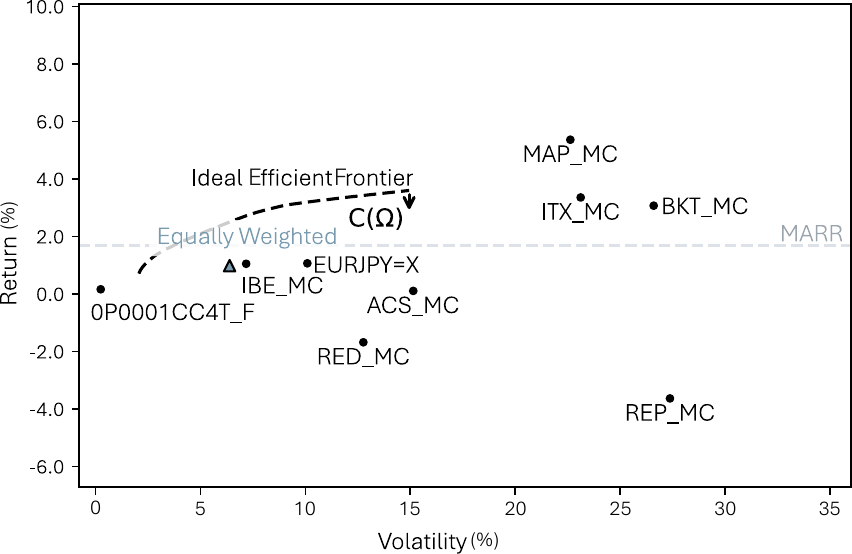}
\caption{Financial performance for the 9-asset portfolio problem studied in Sections~\ref{sec:944} and~\ref{sec:94t} for the $t=4$ time step. The horizontal axis shows portfolio volatility and the vertical axis shows the return. Each black dot marks the volatility-return position outcome of a 100\% allocation to that asset, labeled by its corresponding asset ticker. The cyan triangle represents the equally weighted portfolio, where investment is distributed uniformly across all assets. The dashed black curve represents the \emph{ideal} efficient frontier, which corresponds to the efficient portfolios implied by the assets distribution. The dashed gray line represents the value of the MARR threshold. For reference, the black arrow denoted by $C(\Omega)$ indicates a transaction cost rate of $\nu = 1\%$ -- not applied for these results.
}
\label{fig:EF9_assets_}

\end{figure}

To further frame our financial analysis, we introduce the efficient frontier~\cite{markowits1952portfolio}, which, for a fixed investment window $t$, represents the set of optimal portfolios that minimize risk for a given expected return neglecting transaction costs. In this work, we calculate the efficient frontier using the PyPortfolioOpt Python package~\cite{Martin2021}, respecting all investment restrictions (more details in Appendix~\ref{app:classical_benchmarks}). Note that, since, in this work, we explicitly account for transaction costs, the efficient frontier serves as an \textit{idealized} benchmark for contextualizing our optimized investments, representing a theoretical limit that cannot be fully attained in practice.

Figure~\ref{fig:EF9_assets_} shows the ideal efficient frontier (dashed black curve) for the 9-asset portfolio problem studied in Sections~\ref{sec:944} and~\ref{sec:94t}. Black dots indicate the expected return and volatility for each individual asset over the investment window $t = 4$, while the cyan triangle represents an equally weighted portfolio, where investment is distributed uniformly across all assets. Note that in this work we consider a transaction cost rate of $\nu = 1\%$ which is of the same order of magnitude as the average asset return (indicated by black arrow denoted by $C(\Omega)$). Within this context, an optimized portfolio is considered successful if it lies above the MARR boundary (dashed gray line in the figure) and closer to the efficient frontier than the equally weighted portfolio. In Fig.~\ref{fig:EF9_assets_}, segments of the efficient frontier can lie below the MARR (section with $\text{Volatility} \mathrel{\raisebox{0.2ex}{$\scriptstyle \lesssim$}} 0.25 \times 10^{-2}$ in Fig.~\ref{fig:EF9_assets_}), reflecting realistic investment conditions and highlighting the practical relevance of our optimization.

The efficient frontier is used to frame investment strategies and benchmark portfolio performance. In the next section, we dive into the specifics of the quantum optimization algorithm used to find optimal solutions (investment strategies) for different DPO problems. 

\section{Quantum Optimization Methodology: VQE and ISQR Post-processing}
\label{sec:QuantumSolution}
This section presents the quantum optimization strategy we use in this work to solve the DPO problem. In Subsection~\ref{sec:VQEIntro}, we provide a comprehensive review of the VQE algorithm, highlighting the technical details of the implementation that can contribute to an overall computational speedup. In Subsection~\ref{sec:SQD}, we cover the implementation of the ISQR post-processing technique, an adaptation of the SQD post-processing method~\cite{robledo2024chemistry} tailored to the DPO problem.

\begin{figure}[t]
    \centering
    \includegraphics[width=.475\textwidth]{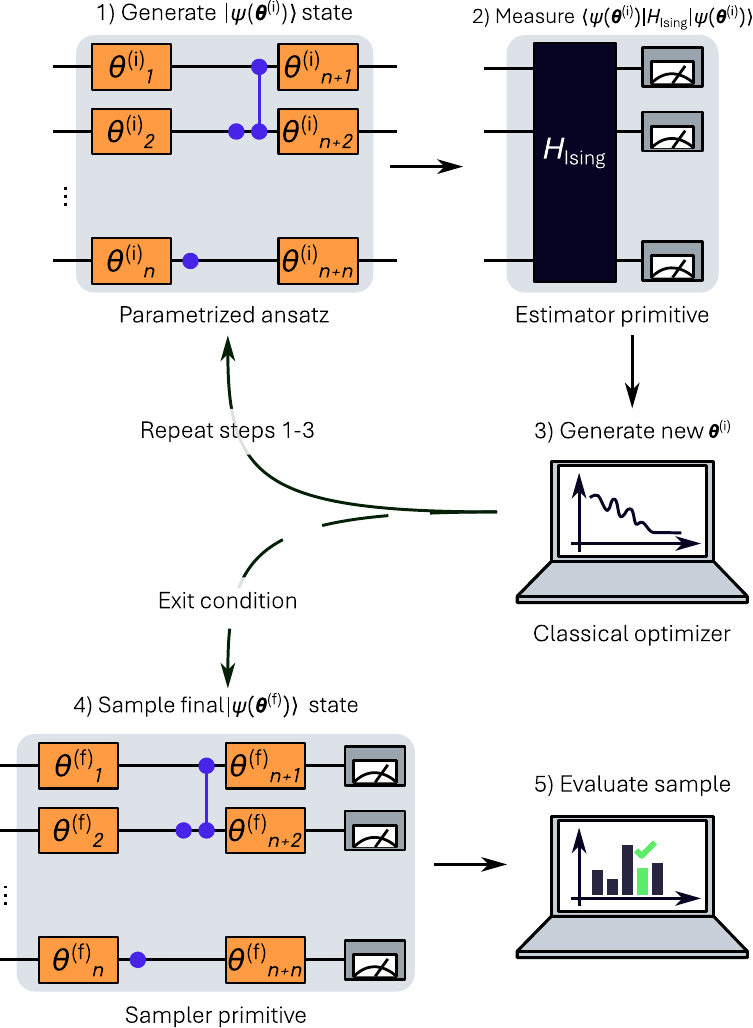}
    \caption{Scheme of the VQE optimization process (described in Section \ref{sec:VQEIntro}). In step 1 we generate an ansatz circuit for a given set of parameters $\bm{\theta}^{(i)}$. In step 2 we measure the expected value of the Ising Hamiltonian $H_\text{Ising}$ encoding the QUBO problem using the estimator primitive of IBM Quantum services. In step 3 we optimize $\bm{\theta}^{(i)}$ with a classical algorithm (DE). If the optimization does not   satisfy the convergence criteria, the process returns to step 1. Once the optimal $\bm{\theta}^{(f)}$ is found, in step 4, we use the sampler primitive from IBM Quantum services to sample the solution. Finally, in step 5 we identify the optimal investment strategy. \\
    Figure adapted from Ref.~\cite{nodar2024scaling}.\label{fig:SchemeVQE}}
\end{figure}

\subsection{Variational Quantum Eigensolver (VQE) \label{sec:VQEIntro}}
In this section, we review the VQE workflow~\cite{peruzzo2014variational,bharti2022noisy,nodar2024scaling} used to solve the QUBO problem introduced in Section~\ref{sec:Formulation} using real QPU devices (specifically, in this work, we use the IBM Fez QPU~\cite{ibm_quantum_fez}). The VQE process consists of optimizing the parameters of an ansatz, such that the resulting circuit minimizes the expected value of an Ising Hamiltonian associated with the QUBO problem. Thus, the states encoded by the ansatz circuit correspond to the optimal solutions to the QUBO problem.

In order to achieve utility scale results in Noisy Intermediate-Scale Quantum (NISQ) devices~\cite{farhi2000quantumcomputationadiabaticevolution,zahedinejad2017combinatorialoptimizationgatemodel,gratsea2024onionvqeoptimizationstrategyground}, identifying and implementing algorithmic improvements, as well as error mitigation strategies, is essential for the efficient operation and scalability of the VQE. In this section, we explore these improvements and provide estimations for the speedups observed during our VQE execution.

The VQE algorithm strongly relies on two key aspects: i) the classical optimizer assisting the VQE algorithm, and ii) the ansatz circuit encoding the quantum state that solves the optimization problem. Regarding the ansatz circuit, we follow the results of our previous work in Ref.~\cite{nodar2024scaling} and deploy an ansatz tailored to the specifications of the DPO problem and the connectivity of the QPU (all details provided in Appendix~\ref{app:ansatzes}). Following the VQE scheme in Fig.~\ref{fig:SchemeVQE}, we introduce (at step 1) the tailored ansatz circuit that generates a $\ket{\psi(\bm{\theta}^{(i)})}$ state, where the circuit parameters $\bm{\theta}^{(i)} \in [-2\pi, 2\pi]$ are updated at each iteration $i$. For the first iteration, $i = 0$, we initialize the parameter vector using random values.

Next, we define an Ising Hamiltonian, ${H}_{\text{Ising}}$, encoding our QUBO problem (see Section~\ref{sec:Formulation}) and evaluate its expectation value, \emph{i. e.}, $\braket{\psi(\bm{\theta}^{(i)})|{H}_{\text{Ising}}|\psi(\bm{\theta}^{(i)})}$ (step 2) in Fig.~\ref{fig:SchemeVQE}. To compute the expected value, we use the \emph{estimator} primitive available on IBM Quantum Runtime~\cite{qiskit_estimator}. The Qiskit Runtime client has undergone significant enhancements compared to our previous work~\cite{nodar2024scaling}, as reported in~\cite{ibm_QDC24}. By enabling the experimental \texttt{gen3-turbo} option, we observed an average speedup of approximately $\times 1.3$ in the quantum evaluations.

Then, we minimize $\braket{\psi(\bm{\theta}^{(i)})|{H}_{\text{Ising}}|\psi(\bm{\theta}^{(i)})}$ by updating the parameters $\bm{\theta}^{(i)}$  with a differential evolution (DE) optimizer~\cite{scipyoptimizedifferential_evolution_nodate, storn1997differential}. We consider two exit conditions for the optimization loop: either the convergence criterion is satisfied (see Appendix~\ref{app:convergence}) or the maximum number of iterations is reached. If neither condition is met, we generate a new set of parameters and repeat the optimization process (steps 1-3 in Fig.~\ref{fig:SchemeVQE}). We define the final set of parameters $\bm{\theta}^{(f)}$ as the one that satisfies either exit condition at iteration $f$.

In order to accelerate the execution of the VQE iterations, we submit quantum jobs to the IBM Quantum Cloud threading the generations of the DE optimizer using the Batch execution mode~\cite{qiskit_batch_api}, resulting in an average speedup of $\sim \times 6$. 

Once the optimization process is finished, we sample the output state $\ket{\psi(\bm{\theta}^{(f)})}$ using the \emph{sampler} primitive~\cite{qiskit_sampler}, which results in
\begin{equation}
        \ket{\psi(\bm{\theta}^{(f)})}_{\text{sampled}} = \sum_{b=1}^{N_s} A_b \ket{b}, 
    \end{equation} 
where $\ket{b}$ is a bit string representing a single investment strategy, and $A_b$ is the quasiprobability of measuring that string. The solution bit string is the one that minimizes $c(b) = \braket{b|\hat{H}|b}$. However, as discussed below in Section~\ref{sec:Efficient_Frontier}, our analysis extends to all sampled bit strings, since the ``best'' investment may change under different financial scoring.

To further accelerate the circuit execution runtime, we reduce the interval between successive shots on the QPU by reducing the \texttt{rep\_delay} option~\cite{ibm2025repetition} from the default $250~\mu s$ to $100~\mu s$. This results in a $\times 2.5$ speedup in every quantum evaluation. Considering all speedups introduced so far: i) \texttt{gen3-turbo}, estimated speedup of $\times 1.3$, ii) DE, estimated speedup of $\times 6$, iii)  \texttt{rep\_delay}, speedup of $\times$2.5, the total performance enhancement that we achieve is around $\times 19.5$. This means that a calculation requiring approximately 8 hours can be completed in only 30 minutes, enabling the application of our method to DPO problems with a large number of assets. We believe that this set of enhancements to the VQE routine can be easily generalized to solve other QUBO problems with large parameter spaces.

\subsection{Ising Sample-based Quantum configuration Recovery (ISQR)}
\label{sec:SQD}
\begin{figure}[t!]
\label{fig:SQD}
\includegraphics[width=0.45\textwidth]{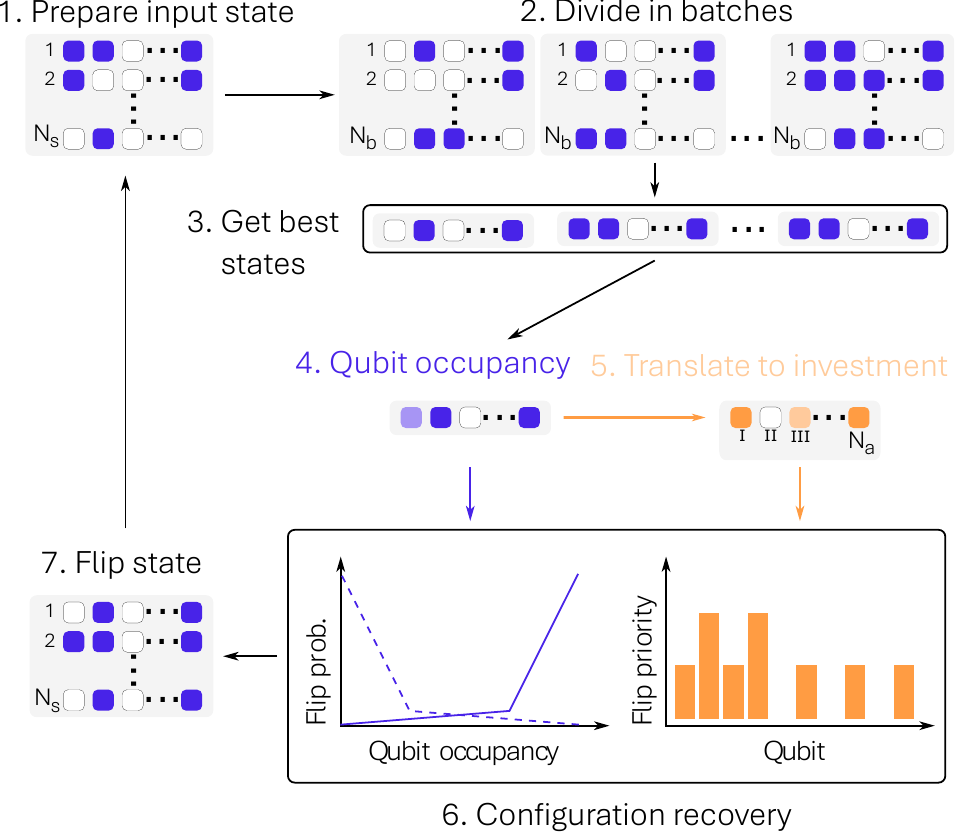}
\caption{Workflow of the ISQR technique applied to the VQE converged solutions. First, a set of bit strings is extracted. After dividing it in batches, the best solution from each batch is studied to determine the bit occupancy pattern. This pattern is then translated into an investment strategy that satisfies the restrictions of the problem. Finally, the CR process iteratively corrects (flips) bits according to these patterns to further refine the solutions.\label{fig:sqd_loop}}
\end{figure}

Originally introduced in Ref.~\cite{robledo2024chemistry}, Sample-based Quantum Diagonalization (SQD) is a noise-aware post-processing technique designed to statistically correct errors on sampled bit strings obtained from quantum hardware. These samples are assumed to approximate the ground state of a target operator, which makes SQD particularly useful (due to its hybrid nature) for simulating large quantum systems, such as large molecules, where exact diagonalization is computationally infeasible.

In this work, we adapt the Configuration Recovery (CR) stage of the SQD to post-process the bit strings resulting from the converged VQE ansatz; by adjusting this stage to the symmetries of the Hamiltonian of our DPO problem. We support this adaptation based on the fact that the Ising Hamiltonian is already diagonal in the chosen basis (Pauli $Z_q$ operators). Thus, we refer to this adapted technique as Ising Sample-based Quantum configuration Recovery (ISQR). 

The ISQR workflow is represented in Fig.~\ref{fig:sqd_loop} as a 7-step iterative loop. In step 1, we extract a set of $N_s$ bit string samples from the quantum device, which serve as candidate solutions encoding sequences of investment decisions across time steps. In step 2, we divide these bit strings into $M$ batches, each covering a subset of $N_b$ bit strings to enable efficient analysis and processing on limited classical hardware. This division into batches allows us to localize the effect of errors in the bit strings due to following the encoding of the DPO problem. 

Next, in step 3, we evaluate the objective function across all selected bit strings within each batch, and, for each batch, we identify the bit string with the lowest (best) cost of the objective function (Eq.~\eqref{eq:QUBO}). With these selected bit strings (one per batch), in step 4, we infer a statistical pattern: the average occupancy of each qubit which corresponds to the average of measuring the state 1 for a given qubit. This pattern captures the dominant structures present in the data and serves as an empirical proxy for the optimal portfolio configuration under noisy sampling conditions.

In step 5, we compute the expected investment profile for the average qubit occupancies using the binary encoding scheme presented in Eq.~\eqref{eq:binary_conversion}. More specifically, at each time step $t$, we compute the total investment weight $K_t$ by summing the investment allocations obtained from the average occupancies. This target reproduces the normalization restriction imposed in the DPO formulation -- Eq.~\eqref{eq:constraint_original_problem} -- serving as a reference for correcting deviations in the original samples. Specifically, if $K_t=1$, the restriction is satisfied; otherwise, $K_t$ is used to prioritize bit-flips in the CR step (below) to enforce the restriction.

At the core of the adapted method lies the CR process, which corrects individual bit strings based on their deviation from the learned investment pattern. In step 6 of the ISQR, we compute the deviation from the corresponding $K_t$, and then determine the minimal number of bit-flips required to enforce compliance for each sample in the original set. This computation exploits the binary encoding: bits associated with higher weights are prioritized, achieving large corrections with fewer flips. Importantly, bit selection is performed by assigning each bit a flipping probability derived from the difference between its current value and the corresponding average occupancy in the learned pattern, modulated by a modified leaky ReLu function and a tunable filling-factor threshold~\cite{robledo2024chemistry}. This probabilistic mechanism favors the reinforcement of statistically dominant configurations while still enabling localized corrections that mitigate noise-induced bit-flips. We refer the reader to Ref.~\cite{robledo2024chemistry} for a detailed discussion on the parameters of the probabilistic bit-flipping routine.

\begin{figure}[t!]
\includegraphics[width=0.45\textwidth]{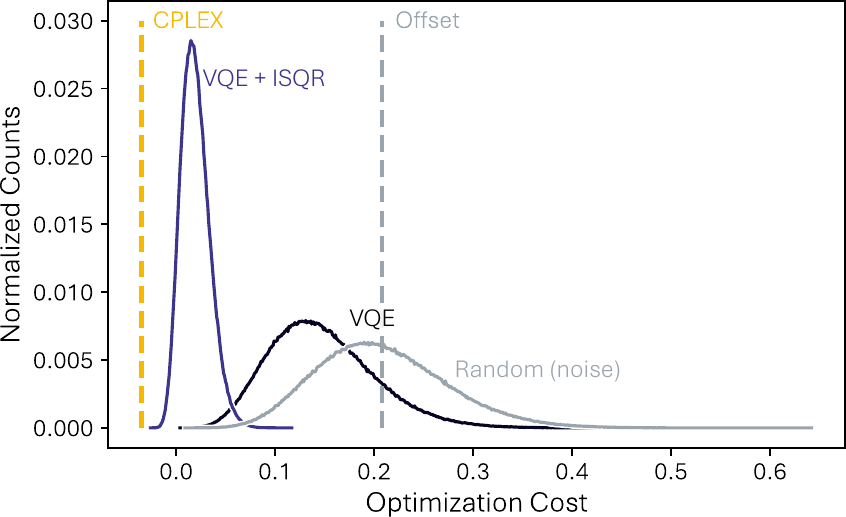}
\caption{Optimization cost distributions of the VQE solutions to the 9-asset DPO problem. To obtain the optimization cost distribution, we account for how many times we found states with the same associated optimization cost. The black curve represents the raw VQE results and the blue curve, the ISQR post-processed results. We add for comparison the random distribution (gray curve), its average (dashed vertical gray line labeled ``Offset''), and the optimization cost obtained by CPLEX solver (dashed vertical yellow line labeled ``CPLEX'').}
\label{fig:944_distribution}
\end{figure}

Last (step 7), in order to generate a new set of corrected bit strings, we flip, according to the computed flip probabilities and priorities, the selected bits in the original samples. These revised configurations align better with both the learned pattern and with the restrictions of the problem, effectively recovering more consistent and reliable solutions. We apply this correction process iteratively, updating the pattern and repeating the evaluation, correction, and refinement steps until we achieve convergence (for two consecutive iterations the change in the optimization cost is less than 2.5\%) -- see Appendix~\ref{app:convergence_ISQR}.

In the following section, we address the 9-asset DPO problem with the VQE approach. We show the results obtained after applying the ISQR technique to post-process our VQE solutions.

\section{VQE and ISQR applied to the 9-asset problem}\label{sec:944}

\begin{table}[t!]
\centering
\begin{tabular}{|l|c|c|}
\hline
\textbf{Optimizer} & \makecell{\textbf{Minimum}\\ \textbf{cost}}& \makecell{\textbf{\% Below}\\ \textbf{offset}} \\
\hline
VQE      & $4.05 \times 10^{-3}$     & 87  \\
VQE + ISQR              & $-2.55 \times 10^{-2}$    & 100 \\
Random (noise)           & $2.63 \times 10^{-2}$    & 53  \\
CPLEX             & $-3.49 \times 10^{-2}$     &  -   \\
\hline
\end{tabular}
\caption{Performance metrics for optimizing the 9-asset DPO across different approaches: minimum optimization costs and, for methods producing cost distributions, the percentage of solutions below the offset.}
\label{tab:optimization_results_944}
\end{table}

\begin{figure*}[t!]
\includegraphics[width=\linewidth]{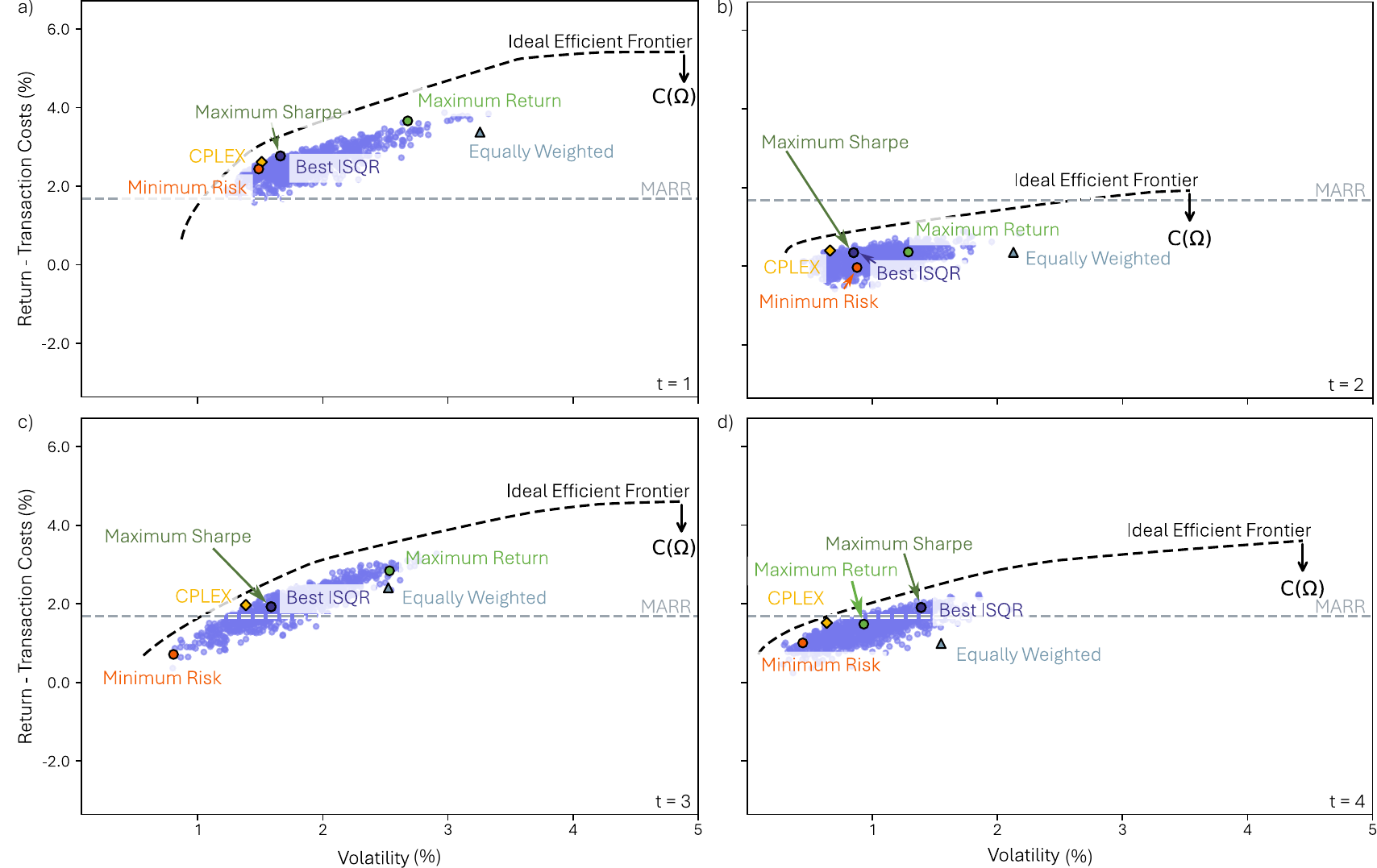}
\caption{Financial analysis for the VQE approach to the 9-asset problem. Panels a), b), c), and d) correspond to $t=1$, $t=2$, $t=3$, and $t=4$ respectively. Each time step is obtained from the full bit string obtained with the VQE. The dashed black curve represents the \emph{ideal} efficient frontier. For reference, the horizontal dashed line represents the MARR of 1.68\% for each time step, the cyan triangle represents the equally weighted portfolio, and the yellow diamond represents the CPLEX solution. The cluster of blue dots represents the 1000 post-processed solutions with lowest optimization cost. From this set, we highlight the following ones: the dark blue dot represents the solution with the lowest objective function (Best ISQR), the orange dot represents the minimum risk investment trajectory, the green dot represents the maximum-return investment strategy, and the dark green dot represents the solution with maximum Sharpe ratio. This solution coincides with the Best ISQR solution for all time steps. }
\label{fig:944_EF_no_assets}
\end{figure*}

\begin{table}[t!]
\centering
\begin{tabular}{|l|c|c|c|}
\hline
\textbf{Solution} & 
\makecell{\textbf{Ann. Sharpe}\\ \textbf{ratio}} &
\makecell{\textbf{Ann. Eff.}\\ \textbf{Return}} &
\makecell{\textbf{Ann.}\\ \textbf{Volatility}} \\
\hline
CPLEX                    & 4.10 & 21.14\% & $4
.73 \%$ \\
Best ISQR                     & 3.74 & 22.62\% & $5.59\%$ \\
Min. Risk               & 2.71 & 13.42\% & $4.32\%$ \\
Max. Return             & 3.51 & 27.16\% & $7.23\%$ \\
Max. Sharpe             & 3.74 & 22.62\% & $5.59\%$ \\
Eq. Weighted            & 2.52 & 23.22\% & $8.53\%$ \\
\hline
\end{tabular}
\caption{Annualized Sharpe ratio, annualized effective return and annualized volatility for each of the solutions highlighted in Fig.~\ref{fig:944_EF_no_assets} for the VQE approach to the 9-asset DPO problem.}
\label{tab:financial_scores_944}
\end{table}

In this section, we apply the VQE workflow and ISQR post-processing routine to a DPO problem with $N_a=9$ assets (see Table~\ref{tab:tickers_param}) across $N_t=4$ time steps, with a resolution of $N_r=4$ qubits. Following the methodology introduced in Ref.~\cite{nodar2024scaling}, we design an ansatz specifically adapted to both the topology of IBM Fez and the structure of the DPO problem (see Appendix~\ref{app:ansatz_9_4_4}). For the classical optimization stage of the VQE, we use the DE algorithm (see Appendix~\ref{app:convergence}).

Figure~\ref{fig:944_distribution} shows the resulting optimization cost distributions for the 9-asset DPO problem under this VQE setup. The left-most dashed vertical yellow line corresponds to the result obtained with the CPLEX optimizer~\cite{docplex} for the QUBO problem. In contrast, the gray curve corresponds to a random distribution of optimization costs, equivalent to sampling from quantum noise. The line labeled ``Offset'' marks the average of the random sampling. We aim to push our results away from this offset and towards optimal cost values (left-side of the figure).

The black curve of the figure corresponds to the optimization costs obtained with the VQE algorithm, without the ISQR post-processing technique. This distribution falls mainly below the offset of the random distribution, implying a significant enhancement of the solution with respect to the random sampling. This finding is consistent with those reported in Ref.~\cite{nodar2024scaling}.

In contrast, the blue curve corresponds to the VQE results after applying the ISQR post processing routine. The distribution becomes significantly narrower and shifted closer towards the minimum optimization cost, demonstrating a substantial boost in solution quality. Notably, the entire distribution lies entirely below the random offset, underscoring the strong effectiveness of the ISQR method in refining quantum outputs. To the best of our knowledge, these are the first results of an SQD-inspired post-processing technique applied to a financial problem.

Table~\ref{tab:optimization_results_944} summarizes this performance under two key metrics: the minimum (best) optimization cost achieved and the percentage of solutions within each distribution that fall below the offset (serving as a signal-to-noise indicator). As we can observe, the percentage below the offset of the VQE results is of $87\%$, indicating a significant increase in consistency of finding a optimal solution, while the ISQR post-processed results yield a distribution entirely below the offset.

To further illustrate the advantages of leveraging the collective set of optimized solutions, we show the financial performance analysis plot in Fig.~\ref{fig:944_EF_no_assets} for the 9-asset DPO problem, with panels a)-–d) representing the four time steps. All panels are derived from the same full bit strings obtained through the VQE, which encode complete investment trajectories. Each bit string is partitioned into $N_t=4$ time-step components to extract the corresponding asset allocations at each time step.

The dashed black line represents the ideal efficient frontier and the dashed gray horizontal line indicates the MARR. As discussed in Section~\ref{sec:Efficient_Frontier}, solutions above the MARR and closer to the ideal frontier are considered successful. The blue cluster of points shows the 1000 solutions with best (lowest) optimization cost after applying the ISQR post-processing to the VQE results. These solutions lie close to the efficient frontier, indicating an overall strong financial performance (see Section~\ref{sec:Efficient_Frontier}). This demonstrates the ability of our approach to generate a diverse set of high-quality portfolios that approximate the optimal risk-return trade-off.

From these 1000 solutions, we highlight four solutions corresponding to the best performance under different scores: i) the lowest optimization cost (Best ISQR, dark blue dot), ii) minimum risk (orange dot), iii) maximum Sharpe ratio (dark green dot), and iv) maximum return (green dot). For comparison, we include in the figure the CPLEX solution (yellow diamond), and an equally weighted portfolio where the investment is distributed uniformly across all assets (cyan triangle).

Table~\ref{tab:financial_scores_944} reports the annualized financial metrics corresponding to each investment highlighted in Fig.~\ref{fig:944_EF_no_assets}. The CPLEX solution provides a solid benchmark, offering a strong trade-off between return and volatility, and achieving the highest annualized Sharpe ratio in the set.

The ISQR-derived solutions offer a broad and informative perspective on the investment landscape. They span a wide range of risk-–return profiles, enabling a detailed characterization of the full set of viable choices available to investors. The best ISQR solution (\textit{i.e.}, the one with minimum optimization cost) achieves a Sharpe ratio comparable to that of CPLEX while delivering an effective return that surpasses the MARR target threshold of $22.04\%$--twice the historical average return of the S\&P 500 (see Section~\ref{sec:Efficient_Frontier}). Additionally, the ISQR solution that maximizes return also demonstrates strong financial performance: it exceeds the MARR threshold and maintains a Sharpe ratio close to that of CPLEX, even though it operates at a higher risk level.

At the lower end of the performance spectrum, the minimum-risk portfolio (from the ISQR solutions) achieves the lowest volatility but at the cost of severely reduced returns, resulting in a smaller Sharpe ratio. At the opposite extreme, the equally weighted portfolio attains an effective return above the MARR threshold--primarily because it incurs no transaction costs, as it is never rebalanced--but it exhibits the highest volatility among all alternatives, making it unnecessarily risky relative to its performance.

Up to this point, we have implemented the ISQR to improve the consistency of the VQE solutions for a 9-asset DPO problem. The resulting optimization costs are significantly pushed towards the CPLEX optimization cost value, which is our classical benchmark. Furthermore, we show the ISQR results on the efficient frontier, surpassing the classical optimizer for certain time steps. However, this VQE workflow imposes the total number of assets across all the time steps to be encoded simultaneously on the number of qubits of the QPU, strongly limiting the size of the portfolios we can work with. In order to solve problems with larger industrial impact, in Section~\ref{sec:ExploringVQEC}, we work with a variation of the VQE algorithm introduced in Ref.~\cite{mugel2022dynamic}, which enables scaling to portfolios with up to 38 assets, maintaining 4 qubits of resolution and 4 time steps.

\section{Exploring the limits of VQE for DPO}
\label{sec:ExploringVQEC}

\begin{figure*}[t!]
\includegraphics[width=\linewidth]{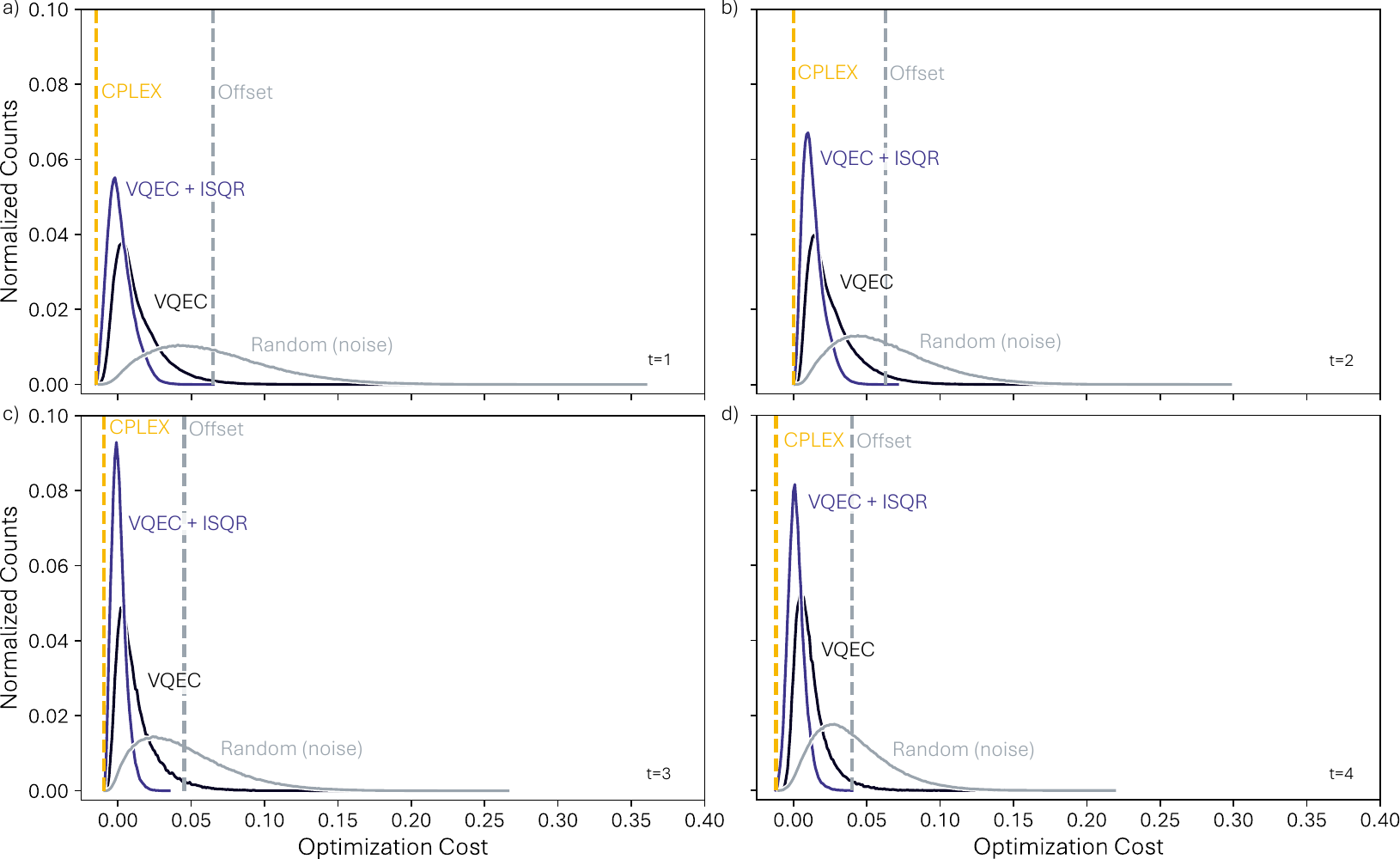}
\caption{Optimization cost distributions of the VQEC solutions for the 9-asset DPO problem for a) time step $t=1$, b) $t=2$, c) $t=3$, d) $t=4$. To obtain the optimization cost distribution, we account for how many times we found states with the same associated optimization cost. The black curve represents the raw VQEC results and the blue curve, the ISQR post-processed results. We add for comparison the random distribution (gray curve), its average (dashed vertical gray line labeled ``Offset''), and the optimization cost obtained by CPLEX solver (dashed vertical yellow line labeled ``CPLEX'').\label{fig:94t_distribution}
}
\end{figure*}

So far, we have deployed an enhanced VQE approach incorporating several key innovations: a speedup over the general VQE method (see Section~\ref{sec:VQEIntro}), an ansatz circuit tailored to both the DPO problem and the QPU symmetries (see Appendix~\ref{app:ansatzes}), and the ISQR noise-aware post-processing technique (see Section~\ref{sec:SQD}). These enhancements allow us to efficiently tackle a 9-asset optimization problem across four different time steps. The deployed method yields optimization quality on par with the classical optimizer CPLEX, and in our specific case, we demonstrate a better financial scoring (see Section~\ref{sec:944}).

However, the implementation of the VQE to solve the DPO problem requires encoding the total number of variables of the problem simultaneously on the available qubits of the QPU, which strongly limits the number of assets, time steps, and bits of resolution that can be represented. To overcome this limitation,  we implement the VQEC, a variation of the VQE algorithm originally proposed in Ref.~\cite{mugel2022dynamic} for DPO problems. The VQEC divides the problem into smaller DPO problems for each individual time step. Specifically, we solve the QUBO objective function for a single $t$ time step on the QPU, and then use the resulting optimal investment $\omega_{t,a}$ as the initial condition for the transaction costs in the next time step. Note that this approach does not allow for the simultaneous exploration of strategies that reduce transaction costs across the full portfolio.



Building upon the formulation of the DPO problem in Section~\ref{sec:Formulation}, it is straightforward to write the constrained variation of a single-time QUBO problem. By factoring out all time-dependent terms in Eqs.~\eqref{eq:return},~\eqref{eq:risk},~\eqref{eq:transaction_costs}, and~\eqref{eq:constraint_gamma}, we can write single-time return ($F_t$), risk ($R_t$), transaction cost ($C_t$), and constraint ($\Gamma_t)$. Consequently, the single-time objective function problem $O_t$ can be written as:
\begin{align}
    O_t =& -F_t+R_t+C_t+\Gamma_t\nonumber\\
    =&-\sum_{a=1}^{N_a}\mu_{t,a} \omega_{t,a} 
    + \frac{\gamma}{2} \sum_{a,b=1}^{N_a}\omega_{t,a} \Sigma_{t,a,b} \omega_{t,b} 
    \nonumber\\+&\!\sum_{a=1}^{N_a}\!\nu_a\lambda_{t,a}\left(\omega_{t, a} -\varphi_{t,a}\omega_{t-1,a}\right)^2 \!+\rho \left( \sum_{a=1}^{N_a}\!\omega_{t,a} - 1\!\right)^{2}\!.
\end{align}
Note that here the single-time transaction cost involves the $\omega_{t-1,a}$ term, which is obtained from solving the previous time step (we remind that in this work we set $\bm{\omega}_{0}=\bm{0}$).

Within this new framework, we solve each time step separately, which significantly reduces the number of required qubits to $N_q = N_a \times N_r$. In this work, we use the IBM Fez QPU, which has 156 qubits, and with this approach, we can fit portfolios with up to $N_a = 38$ assets while maintaining $N_r = 4$ qubits for resolution (see Appendix~\ref{app:ansatz_38_4_t}). We also note that in this formulation, the resulting QUBO matrix for each time step is fully connected (every variable interacts with every other variable), increasing the complexity of the problem.

In the next subsection, we apply the VQEC approach to the same 9-asset problem solved in Section~\ref{sec:944}, allowing for a direct comparison with the VQE results (Section~\ref{sec:94t}). Then, we extend the analysis to an impactful scenario involving 38 assets (Section~\ref{sec:384t}).

\subsection{Extending the 9-Asset Portfolio Problem to VQEC \label{sec:94t}}

\begin{table}[t!]
\centering
\begin{tabular}{|l|c|c|}
\hline
\textbf{Optimizer} & \makecell{\textbf{Minimum}\\ \textbf{cost}}& \makecell{\textbf{\% Below}\\ \textbf{offset}} \\
\hline
 VQEC      & $-3.15\times 10^{-2}$     & 97  \\
VQEC + ISQR              & $-3.59\times 10^{-2}$    & 100 \\
Random (noise)           & $2.63 \times 10^{-2}$   & 53  \\
CPLEX             & $-3.49 \times 10^{-2}$     &  -   \\
\hline
\end{tabular}
\caption{Performance metrics for solving the 9-asset DPO with the constrained method across different approaches, showing minimum optimization costs and, for methods generating cost distributions, the percentage of solutions below the offset. All values represent averages over the four time steps to provide an overall view of the problem.
\label{tab:optimization_results_94t_overall}}
\end{table}

\begin{table}[t!]
\centering
\begin{tabular}{|l|c|c|c|}
\hline
\textbf{Solution} & \makecell{\textbf{Ann. Sharpe}\\ \textbf{ratio}}  
& \makecell{\textbf{Ann. Eff.}\\ \textbf{Return}} 
& \makecell{\textbf{Ann.}\\ \textbf{Volatility}} \\
\hline
CPLEX                    & 4.16  & 22.65\%  & $5.00\%$ \\
Best ISQR                & 4.18  & 22.13\%  & $4.88\%$ \\
Eq. Weighted             & 2.52  & 23.22\%  & $8.53\%$ \\
\hline
\end{tabular}
\caption{Annualized Sharpe ratio, annualized effective return and annualized volatility for each of the solutions highlighted in Fig.~\ref{fig:94t_EF_no_assets} for the VQEC approach to the 9-asset DPO problem.}
\label{tab:financial_scores_94t}
\end{table} 

As a first approach, we demonstrate the VQEC workflow applied to the 9-asset DPO problem with 4 time steps, and benchmark it against the previous VQE approach to solve the same problem in Section~\ref{sec:944}. For the VQEC implementation, we propose a new ansatz tailored to the VQEC method and the connectivity of the QPU (see Appendix~\ref{app:ansatz_9_4_t}).

\begin{figure*}[t!]
\includegraphics[width=\linewidth]{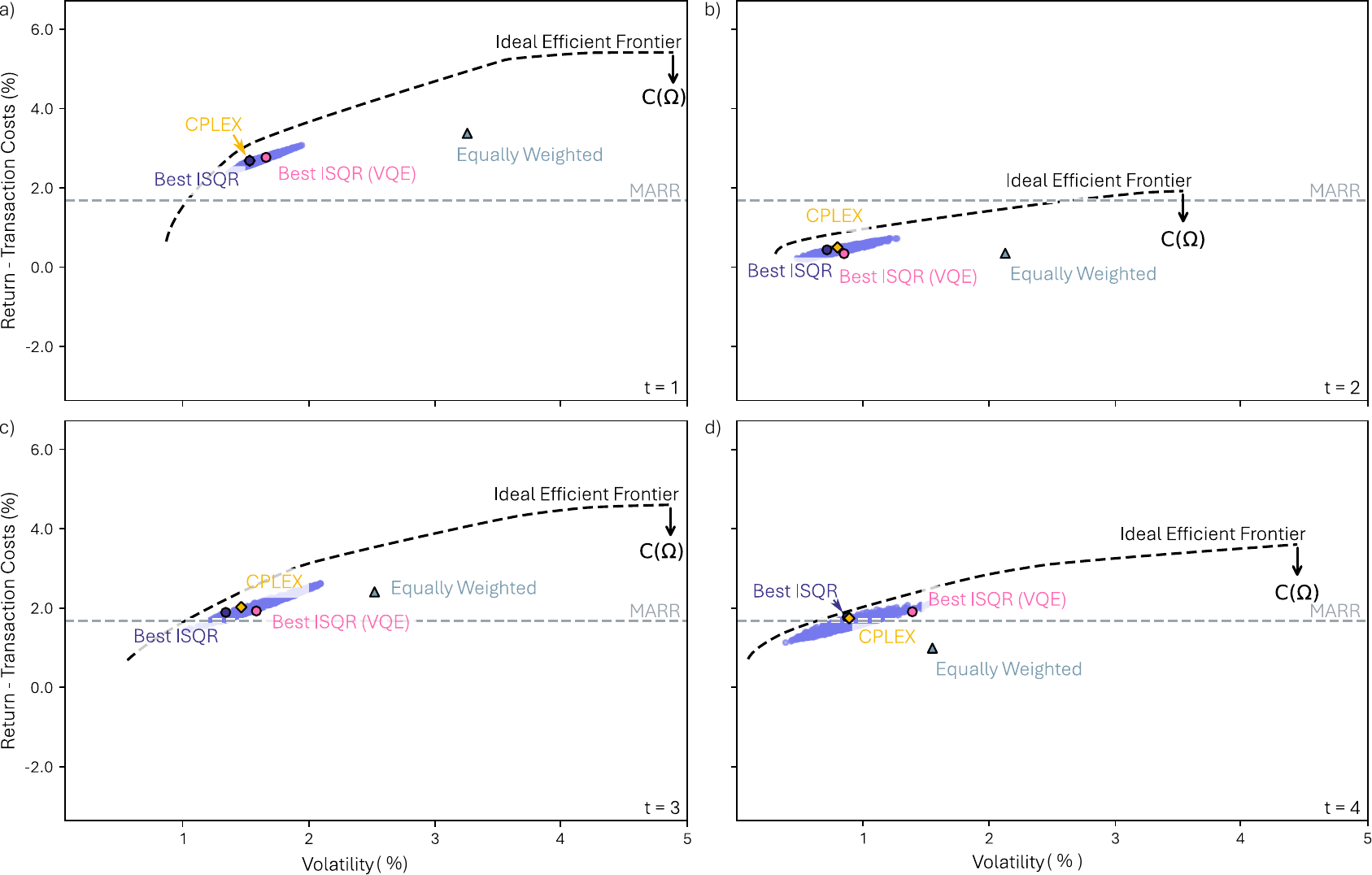}
\caption{Financial analysis for the VQEC approach to the 9-asset problem. Panels a), b), c), and d) correspond to $t=1$, $t=2$, $t=3$, and $t=4$ respectively. The dashed black curve represents the \emph{ideal} efficient frontier. For reference, the horizontal dashed line represents the MARR of 1.68\% for each time step, the cyan triangle represents the equally weighted portfolio, and the yellow diamond represents the CPLEX solution. The cluster of blue dots represents the 1000 post-processed solutions with lowest optimization cost. From this set, we highlight the lowest objective function with a dark blue dot (Best ISQR). Each time step is solved independently
through its corresponding VQEC subproblem. For comparison, we include the Best ISQR (VQE) solution obtained for the same problem using the VQE approach, pink dot.}
\label{fig:94t_EF_no_assets}
\end{figure*}

\begin{figure*}[t!]
\includegraphics[width=\linewidth]{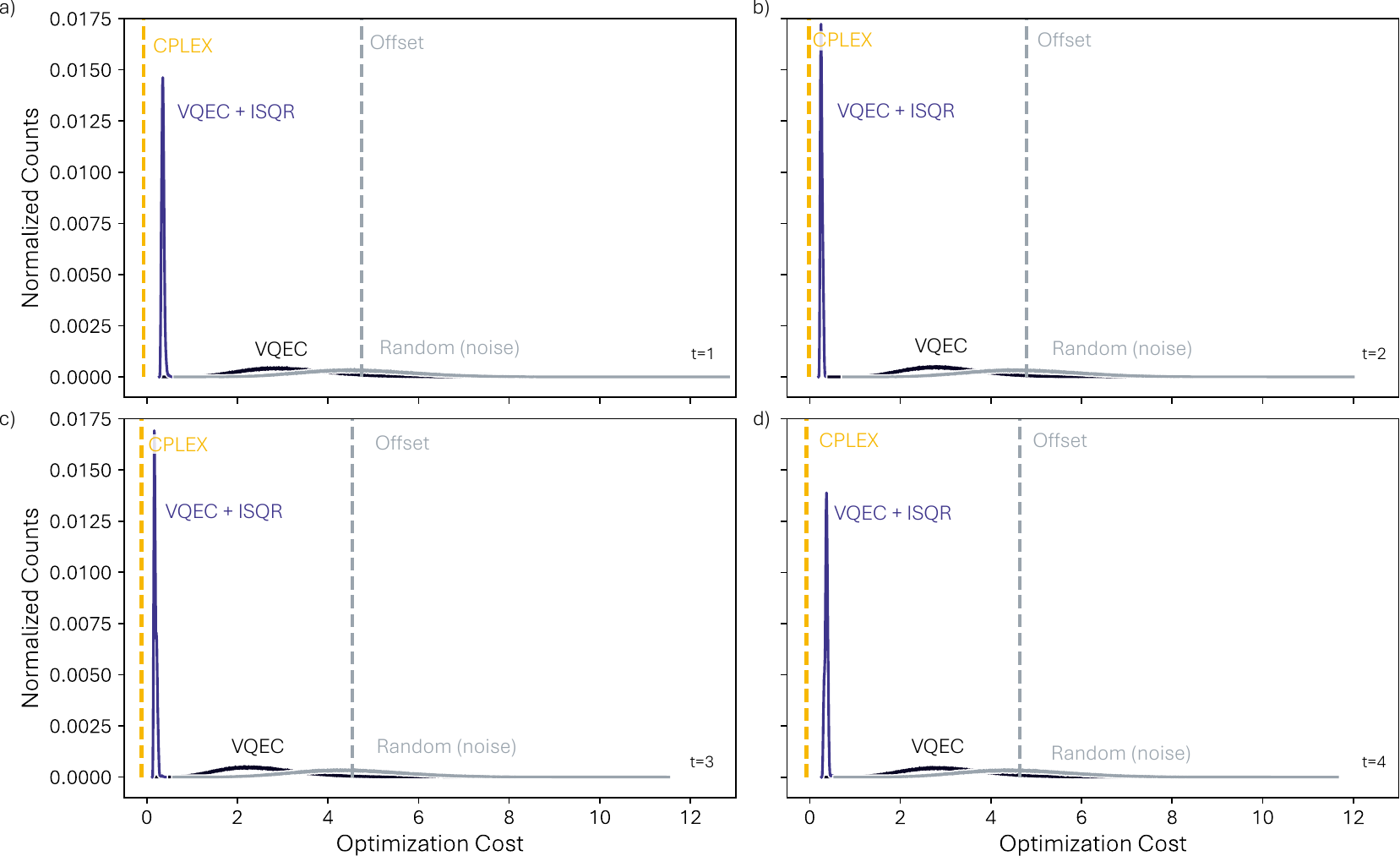}
\caption{Optimization cost distributions of the VQEC solutions for the  38-asset DPO problem for a) time step $t=1$, b) $t=2$, c) $t=3$, d) $t=4$. The black curve represents the raw VQE results and the blue one, the ISQR post-processed ones. We add for comparison the random distribution (gray curve), its average (dashed vertical gray line labeled ``Offset''), and the optimization cost obtained by CPLEX solver (dashed vertical yellow line labeled ``CPLEX'').\label{fig:distribution_384t}}
\end{figure*}

Figure~\ref{fig:94t_distribution}a)-d) shows the resulting distributions of optimization costs for each time considered in the 9-asset DPO problem via the VQEC. As in Fig.~\ref{fig:944_distribution}, we use as reference a random distribution of optimization costs (solid gray curves in Figs.~\ref{fig:94t_distribution}a)-~\ref{fig:94t_distribution}d)). The solid black curves represent the optimization cost distribution of the VQEC optimization workflow, and the blue curve corresponds to the post-processed VQEC results using the ISQR routine. When comparing these results to the full VQE optimization (Fig.~\ref{fig:944_distribution}), we observe that, as in the full VQE case, the VQEC (black curves) yields an optimization cost distribution significantly distinct from the random baseline. These VQEC results indicate the validity and efficacy of the approach for this type of problem decomposition. We also note that the ISQR post-processing technique (blue curves in Figs.~\ref{fig:94t_distribution}a)-~\ref{fig:94t_distribution}d)) still improves the quality and consistency of the VQEC results, although its effect is less pronounced compared to its impact on the full VQE optimization routine (blue curve in Fig.~\ref{fig:944_distribution}).

In Table~\ref{tab:optimization_results_94t_overall}, we show the (average) minimum optimization costs obtained for each method. In particular, we highlight that the VQEC approach, even without the ISQR post-processing, yields a distribution of optimization costs with 97$\%$ of the values below the offset. The ISQR post-processing, as for the VQE approach for the 9-asset DPO problem, pushes all the optimization costs below the offset.

In Figs.~\ref{fig:94t_EF_no_assets}a)-~\ref{fig:94t_EF_no_assets}d), we present the financial performance for each time step of the 9-asset DPO problem solved with the VQEC. In this case, each panel corresponds to an optimal investment obtained only for the corresponding time step. As in Fig.~\ref{fig:944_EF_no_assets}, the blue cluster represents the 1000 lowest optimization cost solutions obtained with the VQEC+ISQR post-processing, and we highlight the best (lowest optimization cost) ISQR solution as a dark blue dot. For comparison, we also include the equally weighted solution (cyan triangle), the previous Best ISQR solution (pink dot) obtained by solving the full problem with the VQE (not VQEC), and the CPLEX solution (yellow diamond) obtained by solving the QUBO problem corresponding to each time step of the VQEC.

Table~\ref{tab:financial_scores_94t} presents the annualized financial metrics for the CPLEX, ISQR, and equally weighted investment strategies. Notably, in this case, the solution obtained with the ISQR post-processing for the VQEC identifies the best-performing investment for the 9-asset problem, over the VQE method in the previous section. It achieves an effective return exceeding the MARR threshold while maintaining very low annualized volatility. This favorable balance results in the highest Sharpe ratio for this problem.

By comparing the financial performance of the VQEC solutions with the corresponding VQE solutions (Section~\ref{sec:944}), we validate the efficacy of our VQEC method for this 9-asset DPO problem: the VQEC method yields a distribution of multiple optimized investments very close to the ideal efficient frontier. In fact, the VQEC solutions are more tightly clustered around the ideal efficient frontier than those from the VQE. This result serves as a proof of concept for using VQEC to solve DPO problems. Next, we apply the VQEC approach to a larger 38-asset DPO problem, leveraging the maximum available qubits on the QPU.

\begin{figure*}[t!]
\includegraphics[width=\linewidth]{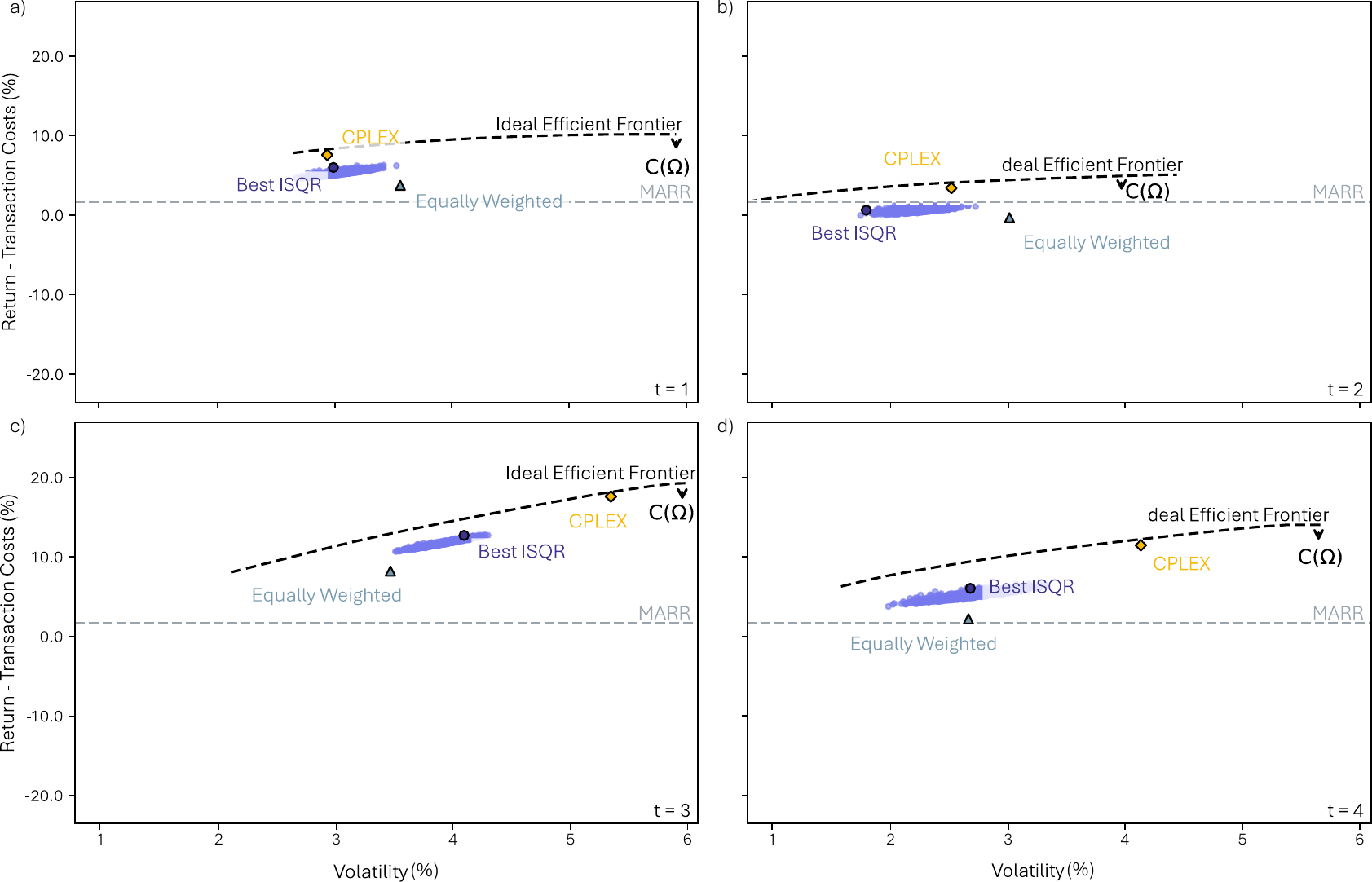}
\caption{Financial analysis for the 38-asset problem via VQEC. Panels a), b), c), and d) correspond to $t=1$, $t=2$, $t=3$, and $t=4$ respectively. The dashed black curve represents the \emph{ideal} efficient frontier. For reference, the horizontal dashed gray line represents the MARR of 1.68\% for each time step, the cyan triangle represents the equally weighted portfolio, and the yellow diamond represents the CPLEX solution. The cluster of blue dots represents the 1000 post-processed solutions with lowest optimization cost. From this set, we highlight the lowest objective function with a dark blue dot (Best ISQR). Each time step is solved independently
through its corresponding VQEC subproblem.\label{fig:EF_384t}}
\end{figure*}

\begin{table}[t!]
\centering
\begin{tabular}{|l|c|c|}
\hline
\textbf{Optimizer} & \makecell{\textbf{Minimum}\\ \textbf{cost}} & \makecell{\textbf{\% Below}\\ \textbf{offset}} \\
\hline
 VQEC      & $0.29$     &  95  \\
VQEC + ISQR              & $0.21$   & 100 \\
Random (noise)           & $0.63$   & 52  \\
CPLEX             & $-0.07$     &   -    \\
\hline
\end{tabular}
\caption{Performance metrics for solving the 38-asset DPO with the constrained method across different approaches, showing minimum optimization costs and, for methods generating cost distributions, the percentage of solutions below the offset. All values represent averages over the four time steps to provide an overall view of the problem.
\label{tab:optimization_results_384t_overall}}
\end{table}

\begin{table}[t!]
\centering
\begin{tabular}{|l|c|c|c|}
\hline
\textbf{Solution} & \makecell{\textbf{Ann. Sharpe}\\ \textbf{ratio}}  
& \makecell{\textbf{Ann. Eff.}\\ \textbf{Return}} 
& \makecell{\textbf{Ann.}\\ \textbf{Volatility}} \\
\hline
CPLEX                    & 9.18  & 130.69\%  & $14.05\%$ \\
Best ISQR                & 7.47  & 82.85\%  & $10.85\%$ \\
Eq. Weighted             & 0.71  & 45.31\%  & $61.10\%$ \\
\hline
\end{tabular}
\caption{Annualized Sharpe ratio, annualized effective return and annualized volatility for each of the solutions highlighted in Fig.~\ref{fig:EF_384t} for the VQEC approach to the 38-asset DPO problem.}
\label{tab:financial_scores_384t}
\end{table}

\subsection{Pushing the limits of VQEC towards applicability: the 38-asset problem\label{sec:384t}}

The example presented in the previous subsection provides initial evidence to support the prospect of the VQEC approach. With this formulation, we now aim to push the existing limits on the DPO problem sizes that can be tackled, which are imposed by the physical properties of the QPU. With this motivation, we increase the number of assets for the portfolio up to $N_a=38$, while keeping the resolution bits fixed to $N_r=4$. Table~\ref{tab:tickers_param} presents the asset tickers along with the asset-dependent transaction fee, $\nu_a$, and the minimum and maximum investment per asset, $m_a$ and $B_a$, respectively, for the 38-asset problem. The four panels in Fig.~\ref{fig:distribution_384t} correspond to the four time steps we study in the 38-asset problem.

We begin by designing a new ansatz tailored to the structure of the problem, using 152 qubits out of the 156 available on the IBM Fez QPU~\cite{ibm_quantum_fez}. The details on the new tailored ansatz are presented in Appendix~\ref{app:ansatz_38_4_t}. 

Figure~\ref{fig:distribution_384t} shows the distribution of optimization costs obtained for each time step for the 38-asset DPO problem for each time step. We include the optimization cost obtained by CPLEX (leftmost dashed yellow line), the random distribution (solid gray line) and its offset (dashed gray line). The black curves represent the optimization cost distribution of the VQEC workflow, and the blue curves correspond to the post-processed solutions using the ISQR routine. The resulting optimization cost distribution is significantly pushed towards the CPLEX value. These results are supported by Table~\ref{tab:optimization_results_384t_overall}, where we include the minimum optimization cost and the percentage of the optimization cost distribution below the offset for each optimization method. We highlight, that, again, the ISQR pushes the distribution below the offset, which is our benchmark for optimal performance. We note that this problem represents a highly complex optimization task, even for state-of-the-art classical solvers. 

In Fig.~\ref{fig:EF_384t}, we show the financial analysis for the four time steps in the problem -- note that the additional assets included in this case allows to reach much larger values of return. The cluster of blue dots in the figure represents the 1000 lowest optimization cost solutions obtained with the VQEC and ISQR post-processing, with the dark blue dot being the best (lowest optimization cost) ISQR solution. For comparison, we include the CPLEX solution (yellow diamond) and the equally weighted solution (cyan triangle). For time steps $t=1, 3, 4$ all the solutions surpass the risk-free rate. Although the best ISQR solution for the 38-asset problem does not outperform the CPLEX benchmark, it still achieves a consistent financial performance, closer to the ideal efficient frontier than the equally weighted portfolio. 

These results are further detailed in Table~\ref{tab:financial_scores_384t}, which reports the financial metrics for the different solutions. Once again, the CPLEX solution establishes a strong benchmark, delivering a remarkable annualized effective return above $100\%$. The best ISQR option provides a strong follow-up, with an $82\%$ annualized effective return and achieving an improved volatility that balances the reduction in return, resulting in a Sharpe ratio comparable to that of CPLEX. In contrast, the equally weighted portfolio, despite yielding a $45\%$ return, remains the worst-performing investment with the lowest risk-adjusted return.

While ISQR slightly underperforms compared to CPLEX in this scenario, it nevertheless produces a competitive and robust solution. Further, there remains room for improvement of the VQEC optimization approach. We note that the current VQEC ansatz circuit (used here and in the 9-asset problem in Section~\ref{sec:94t}) is simply a direct adaptation of the VQE ansatz (in Appendix~\ref{app:ansatz_9_4_4}). We expect that customizing the ansatz circuit for the VQEC, specifically with the aim to increase entanglement, could lead to stronger performance for the VQEC method.


\section{Conclusion and outlooks}
\label{sec:Conclusion}

In this work, we build upon our previous study (Ref.~\cite{nodar2024scaling}) to demonstrate a complete, scalable methodology based on the VQE that is tailored to the DPO problem. Here we introduce upgrades in the methodology that allows us to tackle large-scale portfolio problems. Specifically, we reach a size of 38 assets (including the complete Spanish stock index, IBEX 35)--which is directly impactful for the financial industry. Our methodology incorporates two key innovations: first, the ISQR routine (our adaptation of the configuration recovery subroutine of the SQD technique for DPO), which significantly enhances the consistency and quality of the quantum solution; and second, the VQEC approach, a time-division variation of VQE that tackles the DPO problem step-by-step. By leveraging the VQEC workflow to decompose the problem, we were able to overcome the size and topology limitations of the QPU and study a real-world size DPO problem.

We find that integrating the ISQR routine into the VQE and VQEC workflows significantly improves the probability of finding optimal investment strategies. This consistent performance yields optimized results that are comparable to those obtained by the CPLEX classical optimizer. Furthermore, the quantum approach provides a major advantage: the diversity of optimized investments returned by ISQR. This feature allows us to obtain a large set of financially feasible optimized investments that can outperform CPLEX results under different financial scoring criteria (as demonstrated for the problem discussed in Section~\ref{sec:944}).

In addition, the VQEC approach demonstrates a viable strategy for overcoming the limitations imposed by the number of available qubits, allowing us to successfully optimize a 38-asset DPO problem (as detailed in Section~\ref{sec:384t}). While the current demonstration is robust, we acknowledge that there is still room for improvement. We believe that further methodological optimization of the ansatz circuit design can fully exploit the potential of the VQEC workflow and allow this approach to deliver the performance potential previously shown by VQE, even for large DPO problems.

In summary, our methodology successfully integrates a scalable VQE approach~\cite{nodar2024scaling, mugel2022dynamic} with an adaptation of the configuration recovery technique of the SQD method~\cite{robledo2024chemistry} to create a noise-aware post-processing technique for addressing large-scale financial problems. Our findings show that, even given the current limitations of quantum processors, this tailored approach can provide financially valuable insights, establishing a strong foundation for advantage in quantum optimization.


\section*{Acknowledgments}
This work has been conducted within the quantum ecosystem of Bizkaia, under the BIQAIN (Bizkaia Quantum Advanced Industries) initiative, using infrastructures provided by the Government of Bizkaia.

The authors thank D. Subires and P. Rivero for their valuable input and contributions to this work.

D. A. acknowledges the partial funding of his doctoral research at the University of Deusto, within the D4K (Deusto for Knowledge) team on applied artificial intelligence and quantum computing technologies, thanks to the support of the Basque Government.
\appendix
\section*{Appendices}
\section{Selected Asset Tickers\label{app:data_prep}}
In this work, we optimize the investment strategy for a 9-asset portfolio and for a 38-asset portfolio. The 9-asset portfolio is a subset of the full 38-asset set used in the larger DPO problem. Table~\ref{tab:tickers_param} lists the tickers of all selected assets along with the asset-dependent transaction fee, $\nu_a$, and the minimum and maximum investment per asset, $m_a$ and $B_a$, respectively, associated with each configuration. Although the same 9 assets are used in both problems, the associated parameters differ between the two cases. The Lagrange multiplier $\rho$ in Eq.~\eqref{eq:constraint_gamma} is set to 0.05 for the 9-asset problem and to 0.5 for the 38-asset problem. For the risk aversion coefficient in Eq.~\eqref{eq:risk}, we use $\gamma = 2500$ for 9-asset problem, and $\gamma = 1000$ for the 38-asset problem.

The total set of 38 assets used in this work consists of the Spanish IBEX-35, along with one investment fund and two currency exchange instruments. The selected time period is from 1st November 2022 to 20th February 2023, since, during this period, the IBEX-35 remained compositionally stable for most of the interval~\cite{IBEX35_BME2025}.\\

\begin{table}[!t]
\centering
\begin{tabular}{|c|ccc|ccc|}
\hline
\multirow{2}{*}{\textbf{Ticker}} 
& \multicolumn{3}{c|}{\textbf{9-asset DPO}} 
& \multicolumn{3}{c|}{\textbf{38-asset DPO}} \\
\cline{2-7}
& $\nu_a$ & $m_a$ & $B_a$ & $\nu_a$ & $m_a$ & $B_a$ \\
\hline
0P0001CC4T.F & 0    & 0.05 & 0.5 & 0    & 0.05 & 0.5 \\
ACS.MC       & 0.01 & 0    & 0.3 & 0.01 & 0    & 0.2 \\
ITX.MC       & 0.01 & 0    & 0.3 & 0.01 & 0    & 0.2 \\
EURJPY=X     & 0    & 0    & 0.3 & 0    & 0    & 0.2 \\
MAP.MC       & 0.01 & 0    & 0.3 & 0.01 & 0    & 0.2 \\
BKT.MC       & 0.01 & 0    & 0.3 & 0.01 & 0    & 0.2 \\
REP.MC       & 0.01 & 0    & 0.3 & 0.01 & 0    & 0.2 \\
RED.MC       & 0.01 & 0    & 0.3 & 0.01 & 0    & 0.2 \\
IBE.MC       & 0.01 & 0    & 0.3 & 0.01 & 0    & 0.2 \\

\hline

FER.MC   & - & - & - & 0.01 & 0 & 0.2 \\
ELE.MC   & - & - & - & 0.01 & 0 & 0.2 \\
SCYR.MC  & - & - & - & 0.01 & 0 & 0.2 \\
AENA.MC  & - & - & - & 0.01 & 0 & 0.2 \\
AMS.MC   & - & - & - & 0.01 & 0 & 0.2 \\
MEL.MC   & - & - & - & 0.01 & 0 & 0.2 \\
IAG.MC   & - & - & - & 0.01 & 0 & 0.2 \\
FDR.MC   & - & - & - & 0.01 & 0 & 0.2 \\
LOG.MC   & - & - & - & 0.01 & 0 & 0.2 \\
IDR.MC   & - & - & - & 0.01 & 0 & 0.2 \\
BBVA.MC  & - & - & - & 0.01 & 0 & 0.2 \\
SAN.MC   & - & - & - & 0.01 & 0 & 0.2 \\
SAB.MC   & - & - & - & 0.01 & 0 & 0.2 \\
CABK.MC  & - & - & - & 0.01 & 0 & 0.2 \\
ROVI.MC  & - & - & - & 0.01 & 0 & 0.2 \\
TEF.MC   & - & - & - & 0.01 & 0 & 0.2 \\
NTGY.MC  & - & - & - & 0.01 & 0 & 0.2 \\
ENG.MC   & - & - & - & 0.01 & 0 & 0.2 \\
EURUSD=X & - & - & - & 0 & 0 & 0.2 \\
CLNX.MC  & - & - & - & 0.01 & 0 & 0.2 \\
MTS.MC   & - & - & - & 0.01 & 0 & 0.2 \\
ANA.MC   & - & - & - & 0.01 & 0 & 0.2 \\
ANE.MC   & - & - & - & 0.01 & 0 & 0.2 \\
SLR.MC   & - & - & - & 0.01 & 0 & 0.2 \\
COL.MC   & - & - & - & 0.01 & 0 & 0.2 \\
MRL.MC   & - & - & - & 0.01 & 0 & 0.2 \\
UNI.MC   & - & - & - & 0.01 & 0 & 0.2 \\
GRF.MC   & - & - & - & 0.01 & 0 & 0.2 \\
ACX.MC   & - & - & - & 0.01 & 0 & 0.2 \\

\hline
\end{tabular}
\caption{Tickers with their associated  asset-dependent transaction fee, $\nu_a$, and the minimum and maximum investment per asset, $m_a$ and $B_a$ respectively, for the 9-asset and 38-asset DPO problem. The upper subtable corresponds to the tickers selected for the 9-asset case.  \label{tab:tickers_param}}
\end{table}

\begin{figure*}[t!]
\includegraphics[width=0.9\linewidth]{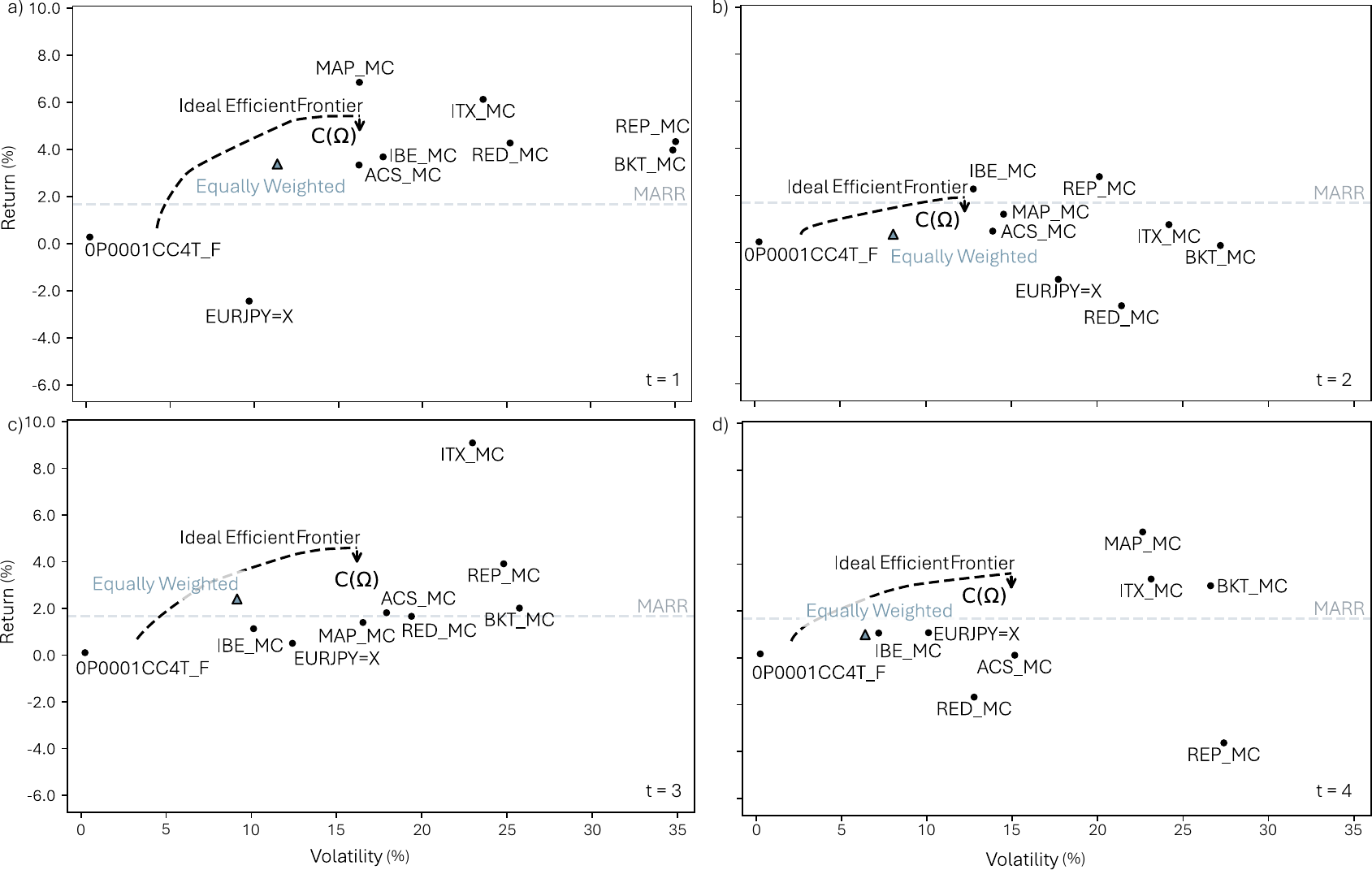}
\caption{Efficient frontiers for the 9-asset problem. Panels a), b), c), and d) correspond to $t=1$, $t=2$, $t=3$, and $t=4$ respectively. The dashed-black curve represents the \textit{ideal} efficient frontier. Each black dot corresponds to investing the entire portfolio in a single asset. The cyan triangle represents the equally weighted portfolio.} 
\label{fig:944_EF_JosefaMadrid_assets}
\end{figure*}

Figure~\ref{fig:944_EF_JosefaMadrid_assets} is a financial performance plot for the 9-asset problem discussed in Sections~\ref{sec:944} and~\ref{sec:94t}. We show the corresponding financial performance of each asset (each one labeled by its ticker) for each time step. For reference, we include the equally weighted portfolio (cyan triangle) and the MARR threshold. 

\section{Details on Formulation}
\label{app:formulation}
In this appendix, we detail the formulation of the asset covariance matrix $\Sigma_t(a,b)$ (Section~\ref{app:subapp:covariance_matrix}), the formulation of the transaction costs term $C(\bm{\Omega})$ (Section~\ref{app:trans_costs}), and the weight normalization routine (Section~\ref{app:normalization}).

\subsection{Asset covariance matrix}
\label{app:subapp:covariance_matrix}

In Eq.~\eqref{eq:risk} of Section~\ref{sec:Formulation}, we introduce the covariance matrix $\Sigma_t(a,b)$ between the assets in the portfolio at the investment time $t$. To calculate $\Sigma_t(a,b)$, we use the \emph{daily} logarithmic return, denoted as $\mu_{s,a}$, for each asset $a$ on day $s$ between consecutive investments -- not to be confused with $\mu_{t,a}$ in Eq.~\eqref{eq:mu_ta}. For this work, we set the time between investments to $\Delta t =28$ days, and we calculate the \emph{daily} logarithmic return as 
\begin{equation}
    \mu_{s, a}=\text{log}\left(\frac{P_{s,a}}{P_{s-1,a}}\right),
    \label{eq:logarithmic_return_days}
\end{equation}
where $P_{s,a}$ represents the closing price of asset $a$ on day $s$. The covariance $\Sigma_t(a,b)$ between assets $a$ and $b$ at time $t$ is then calculated as follows: 
\begin{equation}
    \Sigma_t(a,b)=\frac{1}{\Delta t-1}\sum_{s=(t-1)\cdot\Delta t}^{t\cdot \Delta t}(\mu_{s,a}-\bar{\mu}_{t,a})(\mu_{s,b}-\bar{\mu}_{t,b}),
    \label{eq:covariance_matrix}
\end{equation}
where $\bar{\mu}_{t,a}$ is the average daily logarithmic return of asset $a$ over the same interval.

\subsection{Transaction costs}
\label{app:trans_costs}
Aligned with our previous work in Ref.~\cite{nodar2024scaling}, the original formulation of the transaction costs, $\tilde{C}(\bm{\Omega})$, involves calculating an absolute value,

\begin{equation}
    \label{eq:transaction_costs_app}
    \tilde{C}(\bm{\Omega})= \sum_{t=1}^{N_t } \sum_{a=1}^{N_a} \nu_a\left|\omega_{t, a} -\varphi_{t,a}\omega_{t-1,a}\right|.
\end{equation}

In order to write the DPO problem in the QUBO framework, we need to express the problem depending on quadratic variables. Hence, to write the transaction costs as a positive-defined smooth function of the investment trajectory, we perform the following approximation:
\begin{equation}
    |\omega_{t,a}-\varphi_{t,a}\omega_{t-1,a}|\approx\lambda(\omega_{t,a}-\varphi_{t,a}\omega_{t-1,a})^2,
    \label{eq:ApproxLambda}
\end{equation}
where $\varphi_{t,a}=P_{t,a}/(gP_{t-1,a})$, as defined in the main text (Section~\ref{sec:Formulation}).

The parameter $\lambda$ is introduced to minimize the approximation error in Eq.~\eqref{eq:ApproxLambda}. As detailed in Appendix A of Ref.~\cite{nodar2024scaling}, $\lambda$ is determined by minimizing the average error, $E(\lambda)$, which arises from approximating the absolute value using a quadratic term:
\begin{equation}
E(\lambda) = \int ||x| - \lambda(x)^2| dx,
\end{equation}
where $x = \omega_{t,a}-\varphi_{t,a}\omega_{t-1,a}$. The integral is defined over the range $x\in[0, \max\{\omega_{t,a}-\varphi_{t,a}\omega_{t-1,a}\}]$ (here we are already accounting the positive behavior of $|x|$ and $x^2$). Because the upper integration bound depends on $\varphi_{t,a}$, $\lambda$ must be a piecewise function of $\varphi_{t,a}$:
\begin{equation}
\label{eq:lambda}
\lambda_{t,a} =
\begin{cases}
\sqrt[3]{2}/\left(B_a - \varphi_{t,a} m_a\right), &\quad \varphi_{t,a} \leq 1, \\
\sqrt[3]{2}/\left(\varphi_{t,a} B_a - m_a\right), &\quad \varphi_{t,a} > 1.
\end{cases}
\end{equation}

\subsection{Portfolio weight normalization}\label{app:normalization}
To satisfy the restriction in Eq.~\eqref{eq:constraint_original_problem}, the portfolio weights must be normalized such that their total equals one. A straightforward approach is to divide each weight by the sum of all components. We define $\kappa$ as the sum of all the components (which does not necessarily have to be equal to 1):
\begin{equation}
    \kappa=\sum_{b=1}^{N_a} \omega_{t,b}.
\end{equation}
We can then rescale each $\omega_{t,a}$ as
\begin{equation}
\omega^{\prime}_{t,a} =\frac{\omega_{t,a}}{\kappa},
\end{equation}
so that
\begin{equation}
    \sum_{b=1}^{N_a} \omega^{\prime}_{t,b}=1.
\end{equation}
However, this normalization does not ensure the restriction in Eq.~\eqref{eq:new_constraint} to be satisfied, since the rescaled $\omega_{t,a}'$ weights can be below their lower bounds ($m_a$) or above their upper bounds ($B_a$).

To address this issue, we employ an iterative proportional adjustment that simultaneously enforces both the normalization and bound restrictions. For each time step $t$, the procedure acts on the raw weight vector $\bm{\omega}_t = (\omega_{t,1},\dots,\omega_{t,N_a})$ as follows:

\begin{enumerate}
    \item Initialization and clipping to bounds.  
    Starting from the raw weights, any asset $a$ lying outside its admissible interval $[m_a, B_a]$ is clipped to the nearest bound:
\begin{itemize}
        \item if $\omega_{t,a} < m_a$, set $\omega_{t,a} \leftarrow m_a$;
        \item if $\omega_{t,a} > B_a$, set $\omega_{t,a} \leftarrow B_a$.
    \end{itemize}
    Any asset whose weight is set to a bound ($m_a$ or $B_a$)is marked as \emph{fixed}. The remaining assets are marked as \emph{free}. Let $G$ denote the set of fixed assets and $U$ the set of free ones.

    \item Proportional rescaling of free assets.
    Compute the partial sums for the set of fixed assets, $S_G$, and the partial sum for the set of free assets $S_U$:
    \[
        S_{G} = \sum_{a \in G} \omega_{t,a},
        \qquad 
        S_{U} = \sum_{a \in U} \omega_{t,a}.
    \]
    To satisfy the unit-sum constraint, the free assets must together sum to $S_G + S_U = 1$. As long as $U$ is non-empty, we redistribute this remaining mass proportionally to the current free weights. This is achieved by computing
    \[
        s = \frac{1 - S_G}{S_U},
    \]
    and updating
    \[
        \omega_{t,a} \leftarrow s\,\omega_{t,a},
        \qquad \forall a \in U.
    \]
    \item Update fixed and free sets.  
    After the proportional adjustment, any free asset that now violates its bounds is clipped to the corresponding bound and moved from the set of free assets $U$ to the set of fixed assets $G$. With the updated sets $G$ and $U$, we recompute $S_G$ and $S_U$ and return to Step~2 as long as constraint violations persist.
\end{enumerate}

The procedure continues until all free weights lie strictly within their bounds. Since each iteration fixes at least one additional asset or terminates, the algorithm ends after finitely many steps. The resulting vector~$\bm{\omega}_t$ therefore satisfies both Eq.~\eqref{eq:constraint_original_problem} and Eq.~\eqref{eq:new_constraint} by construction.

\section{Details on classical Benchmarks}
\label{app:classical_benchmarks}
We present the details of the classical benchmarks used in this work. All classical calculations presented in this work, including classical benchmarks and VQE assistance, were performed on a workstation with the following specifications: 32 GB RAM, Intel Core\textsuperscript{TM} i9-13900K CPU, and NVIDIA RTX\textsuperscript{TM} A4000 GPU. 

For reference, we benchmark the quantum results against a randomly generated distribution, which behaves as a worst-case scenario and reproduces a full quantum noise solution. Specifically, we use \texttt{numpy.random} to generate a random distribution of one million bit strings, each representing a random investment strategy. We present such random distributions in Figs.~\ref{fig:944_distribution},~\ref{fig:94t_distribution}, and~\ref{fig:distribution_384t} (gray curves). 

As a benchmark against classical methods, we use the classical solver IBM Decision Optimization CPLEX solver (DOCPLEX)~\cite{docplex}. Here, we limited the execution time to 15 minutes, even if converged is not achieved in that time.

Finally, to plot the ideal efficient frontier introduced in Section~\ref{sec:Efficient_Frontier}, this work uses the PyPortfolioOpt Python package~\cite{Martin2021}. In particular, the \texttt{EfficientFrontier} object from PyPortofolioOpt is instantiated using the covariance matrix computed via Eq.~\eqref{eq:covariance_matrix} and the vector $\mu_t$ obtained from Eq.~\eqref{eq:mu_ta}. The parameter \texttt{ef\_param} is set to \texttt{risk}, which configures the frontier to span a range of risk levels.

\section{Ansatzes for VQE and VQEC algorithms}
\label{app:ansatzes}

In this appendix, we describe the design of the ansatz circuits used to solve the 9-asset and 38-asset DPO problems. We adopt a tailored approach that implements an ansatz adapted both to the QPU (in this work, we exclusively use the IBM Fez QPU) and to the structure of the QUBO problem, as well as to the VQE or VQEC algorithms. Section~\ref{app:ansatz_9_4_4} outlines the general methodology used to design tailored ansatz circuits, in this case, applied for solving the 9-asset problem (with the standard VQE approach), as described in Section~\ref{sec:944}. Section~\ref{app:ansatz_9_4_t} details the modifications introduced to address the same 9-asset problem using the VQEC approach (see Section~\ref{sec:94t}). Finally, Section~\ref{app:ansatz_38_4_t} presents the extension of this strategy to the 38-asset problem solved with the VQEC approach (see Section~\ref{sec:384t}).

\begin{figure}[!t]
    \centering
    \includegraphics[width=.475\textwidth]{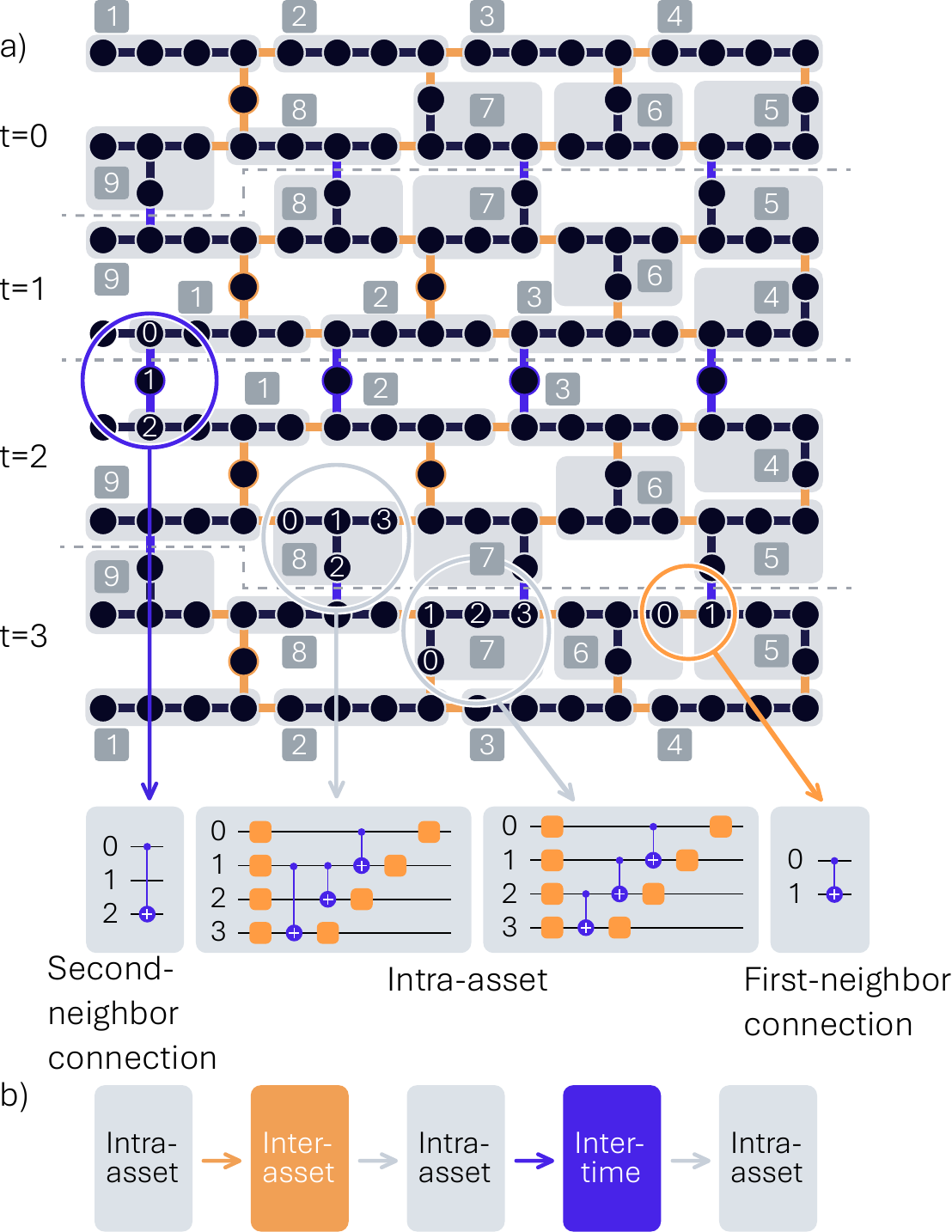}
    \caption{a) Qubit identification of the tailored ansatz of the 9-asset problem from the qubit coupling map of IBM Fez. Gray blocks specify qubit subsets of the QPU identified as an asset at a certain time step. Each block is labeled according to the corresponding asset index. Orange lines represent the interactions between assets at the same time (inter-asset). Blue connections represent the interaction between assets at different times (inter-time). We distinguish between first- and second-neighbor connections. We implement the RA circuit for each gray block.  b) Scheme of layering the different building blocks of the circuit. Each step corresponds to a transformation implemented as a quantum circuit.}
    \label{fig:ansatz_9_4_4}
\end{figure}

\subsection{Ansatz for the VQE approach to the 9-asset problem}
\label{app:ansatz_9_4_4}

The performance of the VQE algorithm strongly depends on the selected ansatz for the quantum optimization stage. Motivated by the results of our previous work in Ref.~\cite{nodar2024scaling}, we adopt the same strategy to build an efficient ansatz to solve the DPO with the VQE algorithm. In particular, we take an intermediate approach in which we simultaneously reproduce the structure of the DPO problem and map it efficiently to the topology of the QPU. 

We implement the ansatz on the IBM Fez QPU -- and we account of the limitations by the number of qubits and physical connections of the QPU. In Fig.~\ref{fig:ansatz_9_4_4}a, we show the mapping of the ansatz for the 9-asset DPO problem onto the IBM Fez qubit map. Each group of four qubits forms a gray block whose resulting quantum state represents the investment over a specific asset at a given time step. The time steps are indicated on the left side of the diagram. A dark gray square labels the index of the asset for each block of four qubits. 

To design an ansatz efficient with current hardware connectivity, we use the Real Amplitudes circuit~\cite{qiskit_RA_2024} as the main building block. This parameterized circuit is implemented on the four-qubit (gray) blocks, which we refer to as the intra-asset circuit. The entanglement pattern depends on the connectivity of each qubit group, as illustrated in Fig.~\ref{fig:ansatz_9_4_4}a. We distinguish two cases: i) for qubits arranged linearly, the 0th qubit is placed on the left-most position, and a reverse-linear entanglement is applied to reach the 3rd qubit; ii) for qubits arranged in a T-shaped structure, the 0th qubit is again placed on the left, but the reverse-linear pattern is modified by replacing the C-NOT between qubits 2 and 3 with a C-NOT between qubits 1 and 3 to suit the T-shaped connectivity (see figure).

In the figure, orange connections represent inter-asset cross terms from the original problem, corresponding in the circuit to C-NOT gates linking a single qubit in one asset (gray) block to a qubit in another asset block within the same time section. The dynamic, inter-time behavior of the DPO problem is captured by the blue connections, which link the four-qubit (gray) blocks of the same asset at different time sections. These blue connections also correspond to C-NOT gates between individual qubits of the same asset across different time steps. Note that all C-NOT connections for both, inter-asset or inter-time, can be implemented directly as first-neighbor interactions or via a second-neighbor interaction using an ancillary qubit, as illustrated in the circuit schemes of Fig.~\ref{fig:ansatz_9_4_4}a).

The division of the qubits into each disjoint gray block allows the sequence of operations to be implemented in parallel. The ansatz circuit scheme for the 9-asset DPO problem is described in Fig.~\ref{fig:ansatz_9_4_4}b): i) we perform all intra-asset circuits, ii) we implement the inter-asset C-NOT gates, iii) we perform another intra-asset circuits layer, iv) we implement the inter-time C-NOT gates, and v) we perform the last intra-asset circuit layer.

\subsection{Ansatz for the 9-asset problem via VQEC}
\label{app:ansatz_9_4_t}

\begin{figure}[!t]
    \centering
    \includegraphics[width=.395\textwidth]{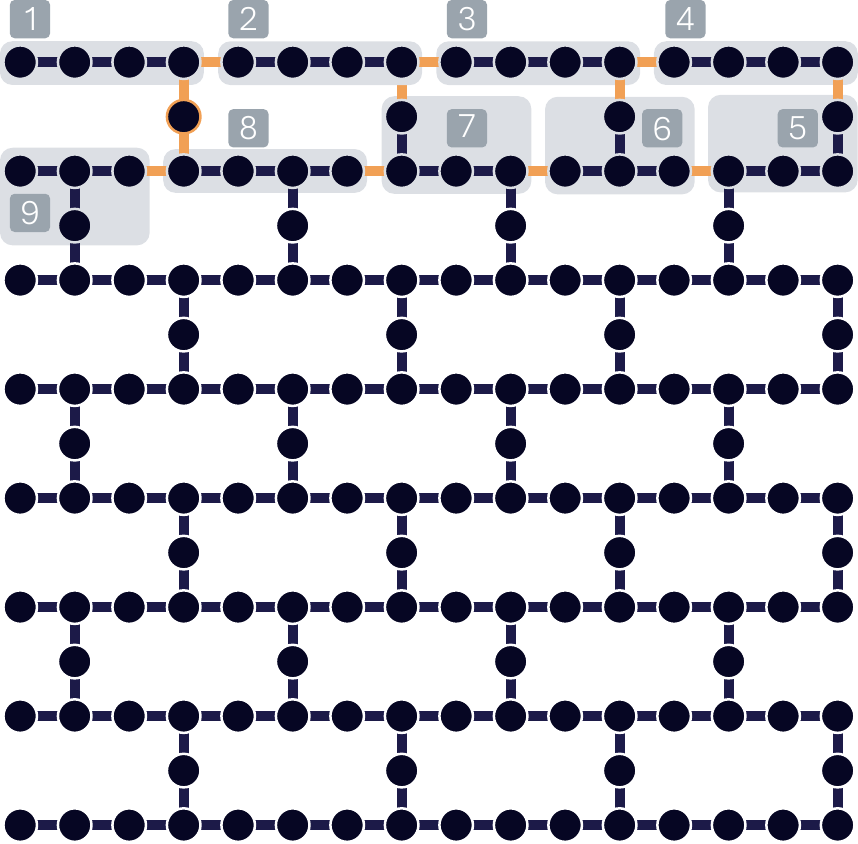}
    \caption{Qubit identification of the tailored ansatz of the 9-asset problem from the coupling map of IBM Fez. Gray blocks specify qubit subsets of the QPU identified as an asset at a certain instant of time. Each gray block is identified by a label, indicating the index of the asset. Orange lines represent the inter-asset connection.}
    \label{fig:ansatz_9_4_t}
\end{figure}

In the VQEC algorithm, each investment rebalance is solved separately. In Fig.~\ref{fig:ansatz_9_4_t}, we show the qubit mapping onto the IBM Fez for the 9-asset DPO problem solved using VQEC in Section~\ref{sec:94t}. The ansatz in this case is a reduced version of the one described in Appendix~\ref{app:ansatz_9_4_4}, corresponding to the asset arrangement for the first time section in Fig.~\ref{fig:ansatz_9_4_4} a). In this ansatz, only inter-asset interactions are represented by the physically connected qubits of the QPU (orange connections in Fig.~\ref{fig:ansatz_9_4_t}).
\begin{figure}[!t]
    \centering
    \includegraphics[width=.395\textwidth]{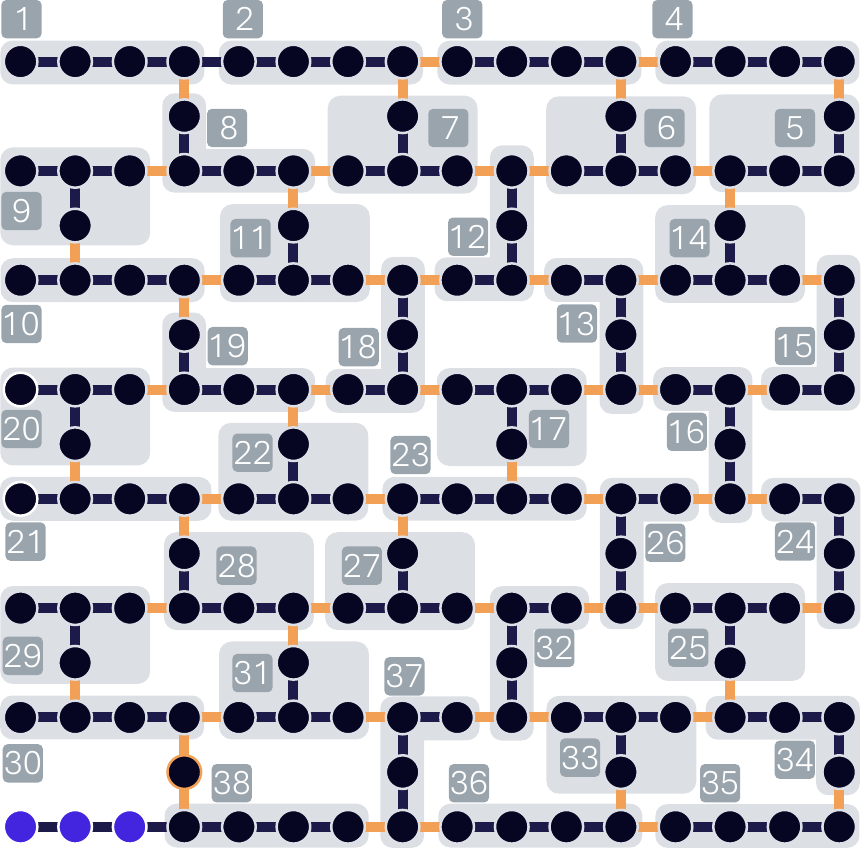}
    \caption{Qubit identification of the tailored ansatz of the 38-asset problem from the coupling map of IBM Fez. Gray blocks specify qubit subsets of the QPU identified as an asset at a certain instant of time. Each gray block is identified by a label, indicating the number of asset. Orange lines represent the inter-asset connection.
    \label{fig:ansatz_38_4_t}}
\end{figure}
This ansatz follows a reduced sequence of operations from the 9-asset DPO problem ansatz presented in the previous section, with the same mapping of the assets as in the $t=1$ section of Fig.~\ref{fig:ansatz_9_4_4}a). In this case, inter-time connections are excluded, which leaves the sequence of operations as follows: i) we perform all intra-asset circuits, ii) we implement inter-asset connections, and iii) we perform the last intra-asset circuit layer. This reduced scheme maintains the essential inter-asset interactions while simplifying the overall circuit design.
\subsection{Ansatz for the 38-asset problem}
\label{app:ansatz_38_4_t}

In Section~\ref{sec:384t} we use VQEC to solve a 38-asset DPO problem. The corresponding ansatz uses 152 out of the 156 available qubits on the IBM Fez QPU. Fig.~\ref{app:ansatz_38_4_t} shows the scheme of the ansatz circuit deployed for this problem, which is designed in a similar way to the ansatz circuit introduced in Appendix~\ref{app:ansatz_9_4_t}: Each four qubits form a gray block that represents an asset (see asset assignment in Fig.~\ref{fig:ansatz_38_4_t}), while orange connections correspond to the inter-asset connections. The sequence of operations of this ansatz is the following:  i) we perform all intra-asset circuits, ii) we implement inter-asset connections, and iii) we perform the last intra-asset circuit layer.

\begin{table*}[t!]
\scalebox{0.9}{
{\renewcommand{\arraystretch}{1.25}
\begin{tabular}{|c|c|c|c|c|}
\hline
\textbf{Problem (Method)} & \textbf{Time-step} & \textbf{Population} & \textbf{Generations} & \makecell{\textbf{Difference in last} \\ \textbf{5 Generations (\%)}} \\ \hline
\textbf{9-asset (VQE)} & All & 188 & 20 & 1.17 \\ \hline
\multirow{4}{*}{\textbf{9-asset (VQEC)}} & t=1 & 28 & 25 & 14.59 \\ \cline{2-5}
 & t=2 & 28 & 25 & 6.30 \\ \cline{2-5}
 & t=3 & 28 & 25 & 10.24 \\ \cline{2-5}
 & t=4 & 28 & 25 & 7.12 \\ \hline
\multirow{4}{*}{\textbf{38-asset (VQEC)}} & t=1 & 112 & 20 & 2.26 \\ \cline{2-5}
 & t=2 & 112 & 20 & 1.15 \\ \cline{2-5}
 & t=3 & 112 & 25 & 3.45 \\ \cline{2-5}
 & t=4 & 112 & 20 & 1.96 \\ \hline
\end{tabular}
}}
\caption{Summary of DE optimization for all DPO problems: population size, number of generations, and difference in mean optimization cost over the last five generations.}
\label{tab:DE_ConvergenceSummary}
\end{table*}

\section{Convergence of the Differential Evolution Optimizer\label{app:convergence}}

\begin{figure}[t!]
    \centering
    \includegraphics[width=.475\textwidth]{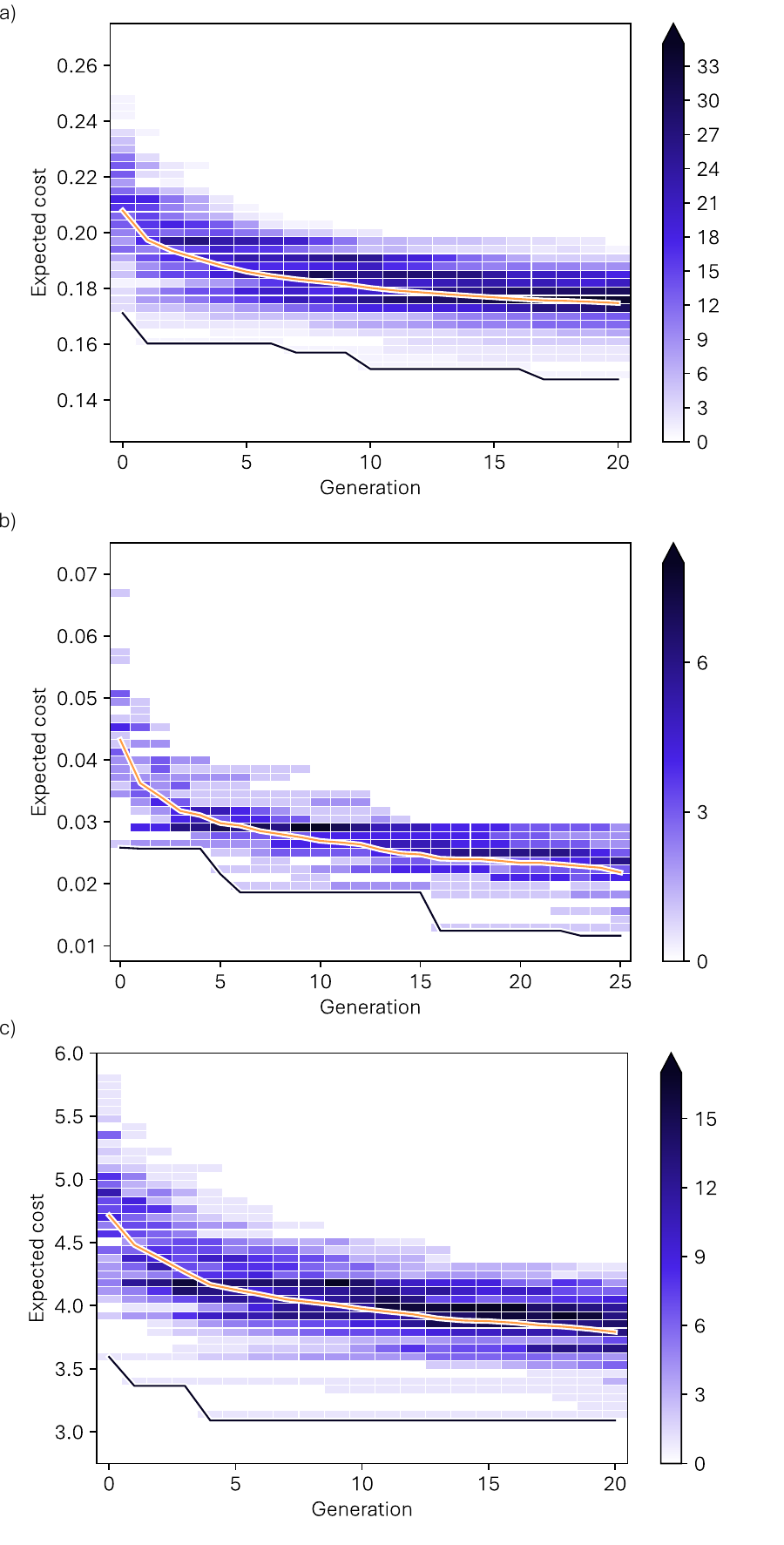}
    \caption{Convergence plots of the a) VQE approach to the 9-asset problem, b) VQEC approach to the 9-asset problem at $t=4$, and c) VQEC approach to the 38-asset problem at $t=1$. Color intensity represents the frequency of parameter sets within each expected cost interval; the orange line shows the average expected cost, and the black line the minimum cost per generation.}
    \label{fig:convergences}
\end{figure}

Classical parameter optimization is performed using a Differential Evolution (DE) algorithm from SciPy~\cite{scipyoptimizedifferential_evolution_nodate,storn1997differential,carrascal2024differential}. By default, each optimization runs for $i=20$ generations, and we consider that convergence is achieved if the difference in mean optimization cost, $\Delta \bar{O} = |\bar{O}[i] - \bar{O}[i-5]|/ \bar{O}[i]$, between the last five generations is $\leq 2.5\%$. If this criterion is not met, we extend the optimization by 5 additional generations, with a maximum of $i=25$ generations.

Table~\ref{tab:DE_ConvergenceSummary} summarizes the results of the convergence for each problem and method, along with the corresponding DE population sizes and the number of generations used. We highlight that, in the 9-asset VQEC case, the optimization cost approaches $\bar{O}[i] \approx 0$ for the optimal parameters, which makes the convergence criterion difficult to satisfy because the denominator in $\Delta \bar{O}$ becomes very small, thereby amplifying minor fluctuations in the mean cost. With that exception, convergence is achieved for most cases.

Figures~\ref{fig:convergences}a)-c) illustrate the convergence behavior for representative cases of the 9-asset (VQE and VQEC) and 38-asset (VQEC) problems. The vertical axis represents the expected optimization cost for all individuals in a generation, discretized into intervals of 0.02. The color intensity indicates the number of parameter sets in a generation that fall within each cost interval. The orange line shows the average cost per generation, while the black line indicates the minimum cost.

\section{Convergence of the ISQR routine\label{app:convergence_ISQR}}
For the DPO problems studied in this work, we use an ISQR configuration of $N_s = 10^6$ bit strings, divided into $M = 10$ batches of $N_b = 10^5$ bit strings each (see Section~\ref{sec:SQD}). The only exceptions are time steps $t = 2$ and $3$ in the 38-asset VQEC problem, where we use $M = 5$ and $N_b = 2 \times 10^5$, which improves convergence. We consider convergence if for two consecutive iterations the change on the optimization costs is lower than 2.5\%.

\clearpage
\bibliography{References}
\end{document}